\documentclass[fleqn,usenatbib,useAMS]{mnras}

\usepackage{newtxtext,newtxmath}
\usepackage[T1]{fontenc}

\DeclareRobustCommand{\VAN}[3]{#2}
\let\VANthebibliography\thebibliography
\def\thebibliography{\DeclareRobustCommand{\VAN}[3]{##3}\VANthebibliography}

\usepackage{graphicx}	% Including figure files
\usepackage{amsmath}	% Advanced maths commands
\usepackage{subfigure}
\usepackage{float}
\usepackage{graphicx}	% Including figure files
\usepackage{amsmath}	% Advanced maths commands
\usepackage{multicol}        % Multi-column entries in tables
\usepackage{orcidlink} % For ORCID iDs
\usepackage{bm}		% Bold maths symbols, including upright Greek
\usepackage{pdflscape}	% Landscape pages
\usepackage[T1]{fontenc}
\usepackage{ae,aecompl}
\usepackage{newtxtext,newtxmath}
\usepackage{hyperref}
\hypersetup{
    colorlinks=true,
    linkcolor=blue,
    filecolor=blue,      
    urlcolor=blue,
    }

\title[MeerKAT search for persistent emission around 25 FRBs]{A MeerKAT search for persistent radio sources towards twenty-five localised Fast Radio Bursts}

\author[L. L. Mfulwane et al.]{
L.~L. Mfulwane\orcidlink{0000-0002-0823-1423},$^{1}$\thanks{mfulwane6@gmail.com}
J.~O. Chibueze\orcidlink{0000-0002-9875-7436},$^{2,3}$
T.~P. Letsele\orcidlink{0000-0002-9225-1303},$^{1}$
T.~M. Nyambe\orcidlink{0000-0001-6575-4792},$^{1}$
C. Venter\orcidlink{0000-0002-2666-4812},$^{1,4}$
\newauthor
M.~C. Bezuidenhout\orcidlink{0000-0001-9541-7439},$^{5,1}$
B.~W. Stappers\orcidlink{orcid.org/0000-0001-9242-7041},$^{6}$
L.~G. Spitler\orcidlink{0000-0002-3775-8291},$^{7}$
M. Caleb\orcidlink{0000-0002-4079-4648},$^{8,9,10}$
A. Deller\orcidlink{0000-0001-9434-3837},$^{11,8}$
\newauthor
J.~A.~S.~Fortunato,$^{12}$
B. Cornejo\orcidlink{0009-0003-0039-0483},$^{13}$
F. Sch\"ussler\orcidlink{0000-0003-1500-6571},$^{13}$
H. Ashkar\orcidlink{0000-0002-2153-1818},$^{13,14,15}$
F. Bradascio\orcidlink{0000-0002-7750-5256},$^{13,16}$
\newauthor
S. Kalita\orcidlink{0000-0002-3818-6037},$^{17,12}$
A. Kundu\orcidlink{0000-0003-2128-1414},$^{18,1}$
M. Kramer\orcidlink{0000-0002-4175-2271},$^{7,6}$
E.~F. Keane\orcidlink{0000-0002-4553-655X},$^{19}$
A. Weltman$^{12,20}$
\\
$^{1}$Centre for Space Research, North-West University, Private Bag X6001, Potchefstroom 2520, South Africa\\
$^{2}$Department of Mathematical Sciences, University of South Africa, Cnr Christian de Wet Rd and Pioneer Avenue, Florida Park, 1709, Roodepoort, South Africa\\
$^{3}$Department of Physics and Astronomy, Faculty of Physical Sciences, University of Nigeria, Carver Building, 1 University Road, Nsukka 410001, Nigeria\\
$^{4}$National Institute for Theoretical and Computational Sciences (NITheCS), Potchefstroom, South Africa\\
$^{5}$South African Radio Astronomy Observatory, Black River Park, 2
Fir Street, Observatory, Cape Town 7925, South Africa\\
$^{6}$Jodrell Bank Centre for Astrophysics, Department of Physics and Astronomy, The University of Manchester, Oxford road, Manchester, M13 9PL, United Kingdom\\
$^{7}$Max-Planck-Institut f\"ur Radioastronomie, Auf dem H\"ugel 69, D53121 Bonn, Germany\\
$^{8}$ARC Centre of Excellence for Gravitational Wave Discovery (OzGrav), Hawthorn, VIC 3122, Australia\\
$^{9}$CSIRO, Space and Astronomy, PO Box 1130, Bentley WA 6102, Australia\\
$^{10}$Sydney Institute for Astronomy, School of Physics, The University of Sydney, NSW 2006, Australia\\
$^{11}$Centre for Astrophysics and Supercomputing, Swinburne University of Technology, Hawthorn, VIC, 3122, Australia\\
$^{12}$ High Energy Physics, Cosmology and Astrophysics Theory (HEPCAT) Group, Department of Mathematics and Applied Mathematics,\\ University of Cape Town, Rondebosch, Cape Town, 7700, South Africa\\
$^{13}$IRFU, CEA, Universit\'e Paris-Saclay, F-91191 Gif-sur-Yvette, France\\
$^{14}$Laboratoire Leprince-Ringuet, \'Ecole Polytechnique, CNRS, Institut Polytechnique de Paris, F-91128 Palaiseau, France\\
$^{15}$ Institute of Space Sciences (IEEC-CSIC), Campus UAB, Torre C5, 2a planta, 08193 Barcelona, Spain\\
$^{16}$ Universit\'e Paris-Saclay, CNRS/IN2P3, IJCLab, 91405 Orsay, France\\
$^{17}$Astronomical Observatory, University of Warsaw, Al. Ujazdowskie 4, Warsaw 00478, Poland\\
$^{18}$ NASA Postdoctoral Program Fellow, Astrophysics Science Division, NASA Goddard Space Flight Center, Greenbelt, MD 20771, USA\\
$^{19}$School of Physics, Trinity College Dublin, College Green, Dublin 2, D02 PN40, Ireland\\
$^{20}$African Institute for Mathematical Sciences, 6 Melrose Road, Muizenberg, Cape Town, 7945, South Africa\\
}

\date{Accepted XXX. Received YYY; in original form ZZZ}

\pubyear{\the\year{}}

\begin{document}
\label{firstpage}
\pagerange{\pageref{firstpage}--\pageref{lastpage}}
\maketitle

\begin{abstract}
The discovery of persistent radio sources (PRSs) associated with repeating fast radio bursts (FRBs) has shed light on the immediate environments and possible progenitors of these FRBs. The confirmed PRSs may support the theory that FRB progenitors are compact central engines, whilst the non-detections suggest diversity of FRB's local environment. 
We perform a subarcsecond-resolution MeerKAT search at 1.28 GHz on 25 well-localised FRB positions provided by ASKAP and MeerTRAP. We detect 14 radio sources and provide flux upper limits for 12 non-detections (both these numbers include a source that was detected during two epochs of observation, and not detected during one epoch, adding up to 26).
One radio source shows variability as seen in flux variations over three epochs of observation. 
Archival optical data reveal excesses in the direction of 13 detected radio sources. Similarly for four sources in the X-ray band, with one possibly being a high-energy signature of a radio galaxy core.
Since we cannot definitively classify our detected radio sources as PRSs, future high-resolution observations with e-MERLIN will be required to resolve the radio emission and pronounce on the presence of compact PRSs associated with the 14 detected sources presented here.
\end{abstract}


%%%%%%%%%%%%%%%%%%%%%%%%%%%%%%%%%%%%%%%%%%%%%%%%%%

%%%%%%%%%%%%%%%%% BODY OF PAPER %%%%%%%%%%%%%%%%%%

\section{Introduction}
Fast Radio Bursts (FRBs) are extremely bright, extragalactic radio sources that last for sub-milliseconds. The first FRB was discovered by \citet{lorimer2007} in the Murriyang (Parkes) data archive. More than 800 FRBs have been discovered since this first burst, of which 58 have been reported to repeat \citep{chimecollabo, an2025, wang2025}. Therefore, there are two identified FRB populations, one-off FRBs and repeating FRBs. However, it is uncertain whether the majority or even all FRBs repeat on some cadence. Repeating FRBs are observed to produce multiple bursts over a short period of time, exhibiting a variety of burst energies. On the other hand, one-off FRBs are observed to produce only one burst. This suggests diverse emission processes and environments. A recent study by \citet{Kirsten2024} proposed that while the early studies suggest two populations indicate different environments, with longer observation time the burst energy distribution might suggest the possibility of the same environment under different physical conditions. Therefore, it is still not clear if the two population share the common origin.

Upon localisation of repeating FRBs such as FRB20121102A \citep{Chatterjee2017}, FRB20190520B \citep{Niu2022}, FRB20190417A \citep{moroianu2025}, and FRB20240114A \citep{2025Bruni}, compact persistent radio sources (PRSs) have been confirmed to be co-located with these FRBs at the milliarcsecond level. A PRS candidate associated with FRB20201124 was detected by \citet{Ravi_2022} using Very Large Array (VLA) and \citet{2024Bruni} also detected the PRS with VLA observations at 15 and 22~GHz. The PRSs associated with FRB20121102A and FRB20190520B share several notable properties, including high radio luminosities and negative spectral indices, consistent with non-thermal synchrotron emission. Both are localised to dwarf host galaxies and are compact on milliarcsecond scales, suggesting emission from dense, confined regions. Furthermore, each exhibits variability in rotation measure (RM), indicating turbulent and magnetised local environment \citep{Chatterjee2017, Niu2022}. In contrast, the PRS associated with FRB20201124A shows a lower luminosity and a positive spectral index, consistent with a star-forming origin on sub-arcsecond scales \citep{Nimmo_2022, 2024Bruni}. Similarly, the PRS associated with FRB20190417A shows low luminosity and a negative spectral index; it is localised within a star-forming dwarf galaxy and is compact at milliarcsecond scale, also suggesting a confined environment \citep{ibik2024, moroianu2025}. The PRSs exhibit luminosities that range from $L_{\nu} \approx 10^{27} - 10^{29}$ ergs s$^{-1}$ Hz$^{-1}$ and are usually spatially offset from the centre of their host galaxy. \citet{Yang_2020} proposed a relation between the PRS luminosity and the FRB's RM, suggesting that highly luminous PRSs are consistent with larger RMs. The emission from the compact radio source can persist on timescales of days, months, to years. Therefore, PRSs are compact radio sources with a long lifespan that are not active galactic nuclei (AGNi), but brighter than the galaxy's local star formation signature \citep{ibik2024}. However, so far no PRS has been associated with a one-off FRB; therefore, this suggests a diversity of FRB progenitors or different emission mechanisms \citep{an2025}. 

There are several models addressing the FRB progenitor question, most of which are linked with young and rapidly rotating neutron stars (isolated ones or merging ones;  \citealt{Margalit2019}). \citet{totani2013, yamasaki2018} proposed that the progenitor may be a merger between binary neutron stars. \citet{popov2013} proposed hyperflares and giant flares from magnetars, while \citet{Kashiyama2013} suggested white dwarfs as the progenitor. 
A leading theory is a single central engine (magnetar), powering both the FRB and the PRS (magnetar wind nebula). For example, \citet{Margalit2019} proposed persistent emission as a result of a magnetar nebula, suggesting that the emission is powered by the relativistic electrons and magnetic fields energised by magnetar flares and leading to nebular expansion. Another model is the ultra-relativistic pulsar wind nebula that sweeps up its surrounding medium, with FRBs repeatedly produced via several mechanisms \citep{2017Dai}. 

The presence of a PRS coinciding with an FRB offers crucial clues about the PRS progenitor. Probing the immediate environments of PRSs enables us to explore the mechanisms responsible for the persistent emission and better understand whether the associated PRSs are a universal feature of all repeating FRBs or are linked to specific progenitor systems. \citet{law2022} mentioned that PRSs could be associated with one-off FRBs, as the association of FRB / PRS is unclear. Given the probable diversity of FRB environments, a detailed study of several FRBs will be essential to understand their emission mechanisms and progenitor channels. 

\citet{an2025} made an argument that the proposed diversity in FRB progenitors may rather be a result of observational limitations instead of physical intrinsic differences; i.e., one-off FRBs could be associated with faint PRSs which are below the sensitivity threshold. Therefore, there may be an evolving progenitor: the active FRB bursts and luminous PRS require a young and energetic progenitor, while one-off FRB with a fainter PRS requires an evolved, older progenitor. To determine if the hypothesis that one-off FRBs could be generated by evolved progenitors resulting in faint PRS holds, or if PRSs are exclusive to repeating FRBs, it is necessary to conduct an extensive PRS search using a substantial sample of well-localised one-off FRBs. Therefore, to elucidate the nature of FRB progenitors and the characteristics of the environment, and to address the question whether PRSs are solely associated with repeaters or also with one-off FRBs, we aim to search for PRSs associated with one-off FRBs. 

Finding a strong association between an FRB and a PRS is not straightforward. An association is deemed reliable for $\sim 20$ arcseconds localisations \citep{Nimmo_2022, ibik2024}. However, other processes not related to the FRB/PRS association can also generate the persistent radio emission. The most significant among these is star-formation-related radio emission from the host galaxy due to synchrotron emission \citep{ibik2024}. The PRS can be distinguished from that of other persistent radio emission based on its angular size and/or spectral index, but this requires multi-frequency observations as well as observations at different angular scales \citep{2017Marcote,Bhandari_2023,Dong_2024}.

The primary purpose of this work is to provide a finder chart for PRSs using the MeerKAT telescope. We search for PRS candidates that are located near FRB positions that have been well localised by the Australian SKA Pathfinder (ASKAP) telescope and MeerTRAP. Our detected radio flux provides an upper limit for the potential PRS flux (due to possible contamination from the host galaxy emission) and can support high angular resolution and/or high-frequency follow-up observations. In addition, we searched for archival data in the optical and X-ray bands in different catalogs in the direction of our detected radio sources. This paper is structured as follows: Section~\ref{observations} discusses observations and data reduction; Section~\ref{results} describes the radio continuum detection. Section~\ref{discussion} discusses our new source detections, and lastly Section~\ref{conc} presents the conclusions and future work.

\section{Observations and Data Reduction}\label{observations}

\subsection{FRB Sample}

To search for PRSs candidates using the MeerKAT telescope, we used a sample consist of 25 one-off FRB positions, localised by ASKAP/CRAFT and MeerTRAP (see Section~\ref{appendix}). In addition to selecting well-localised sources, the selection criteria of these FRBs were not strict as such. The selected sample exhibited lower DMs and redshifts, increasing the likelihood of detecting PRSs. In addition, radio bursts that had repeater-like properties were also taken into consideration, because only repeaters seem to be associated with PRSs. Our MeerKAT observations have been obtained based on 2021, 2022, and 2023 open time proposals (SCI-20210212-CV-01, SCI-20220822-CV-01, SCI-20230907-CV-01, respectively, see \citealt{2022MNRAS.515.1365C}).

\subsection{MeerKAT Observations}
 The MeerKAT radio telescope is a 64-antenna interferometer array that is located in the northern Karoo desert. Each dish is a Gregorian parabolic surface with an effective diameter of 13.5 metres. Of the 64 dishes, 48 are located within a 1~km radius around the array's inner core, and the remaining 16 are spread outward up to 8~km. The MeerKAT array's smallest and longest baselines are 29~m and 8~km, respectively. The angular scales range from 5~arcseconds to 27~arcminutes at the central frequency of the L-band (1283 MHz). MeerKAT was used to observe different FRB positions with 90 minutes of on-source integration time, and a phase calibrator observed for 2~minutes for every 15~minutes on the targeted FRB position. The observation period is repeated for each epoch of these FRB fields at the L-band (856-1712 MHz). Data were reduced using oxkat3, a semi-automated MeerKAT data analysis pipeline \citep{2020Heywood}.

\subsection{Calibration and Imaging}
The oxkat3 pipeline utilises radio interferometric data reduction software that is publicly available. The pipeline provides reduced and calibrated visibility data, continuum images, and diagnostic plots as the final data. In addition, the customary configuration includes flagging, cross-calibration, and self-calibration processes. In the case of the flagging process, the low-gain bandpass edges, which are 856-880 MHz and 1658-1800 MHz, are flagged on all baselines. The location of the Galactic neutral hydrogen line (1419.8-1421.3 MHz) is also flagged to remove line contamination as a result of the continuum imaging process. The radio frequency interference (RFI) is also flagged out in regions with baselines shorter than 600~m. Other RFI that might have impacted the data are flagged using case tasks such as RFLAG and TFCROP for the calibrators and the TRICOLOR package for the target fields \citep{2007Mcmullin}.

In the case of cross-calibration, oxkat uses casa tasks, which includes standard steps such as setting the flux scale and deriving corrections for residual delay calibration, bandpass, and time-varying gain \citep{2007Mcmullin}. After all corrections are applied on target fields, every five consecutive frequency channels are averaged. Then only the calibrated visibilities from the field target are extracted to obtain the science target. The target data are deconvolved and imaged using the WSClean imager, which comprises of the multiscale and wideband deconvolution algorithms (which allows better imaging of diffuse emission present in our fields). The full bandwidth is divided into 10 subbands, each with a bandwidth of 82~MHz. Then deconvolution is performed in each subband images.

WSClean generates the multi-frequency synthesis (MFS) map \citep{2014Offringa}, which is a full bandwidth map consisting of a central frequency of 1283 MHz in joined-channel deconvolution mode. Each of the 10 subbands is deconvolved with an initial high mask of $20\sigma_{\rm rms}$ using the auto masking function provided by WSClean, where $\sigma_{\rm rms}$ is the root mean square noise. After the deconvolution process, an artefact-free model of the target field is generated for the self-calibration process. In the final iteration of imaging, the masking threshold is reduced to $3\sigma_{\rm rms}$. For self-calibration, the oxkat pipeline uses tasks from the Cubical software \citep{2018Kenyon}. The flux densities reported in this work were measured using the casa task \textit{imfit}\footnote{A detection in a MeerKAT image is defined as a signal with a significance of $\geq 3\sigma_{\rm rms}$. Accordingly, a region is carefully drawn along the inside $3\sigma_{\rm rms}$ contour line of the target source using CASA Viewer to obtain an accurate flux. The CASA \textit{imfit} task is subsequently used to perform a Gaussian fit. This method yields best-fit parameter values, including peak flux density, integrated flux density, position (RA,DEC), and source size (major/minor axes, position angle) of the target source. In addition, the corresponding errors for each fitted parameter are provided in Table \ref{tab:info1}.}.

\begin{figure*}
\centering
    
    \begin{minipage}{1.0\textwidth}
    \centering
        \includegraphics[width=1.0\textwidth]{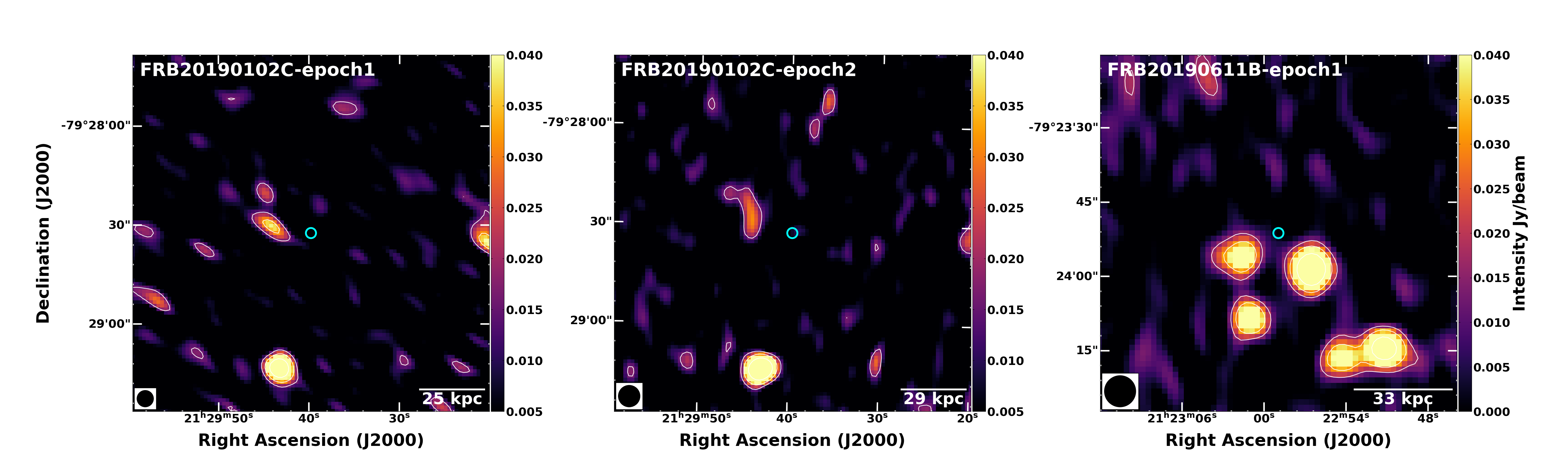}
    \end{minipage}
        
    \begin{minipage}{1.0\textwidth}
    \centering
        \includegraphics[width=1.0\textwidth]{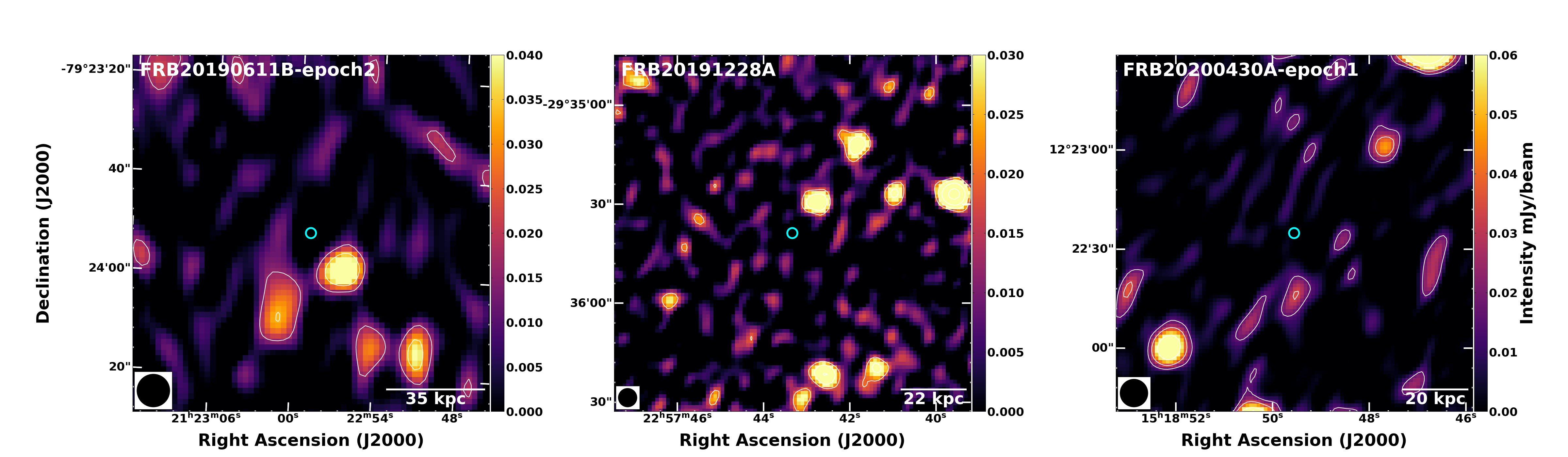}
    \end{minipage}

    \begin{minipage}{1.0\textwidth}
    \centering
      \includegraphics[width=1.0\textwidth]{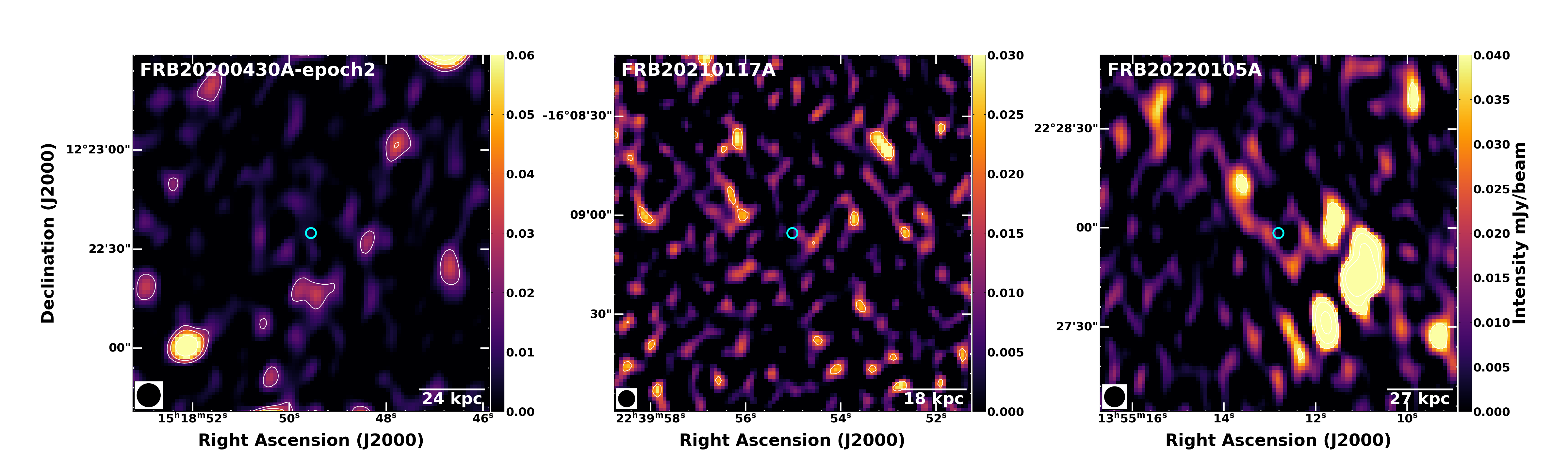}
    \end{minipage}

    \caption{Non-detections by MeerKAT. The cyan circle indicates the position of the ASKAP/MeerTRAP FRB. White contours corresponding to $3,6,12,24$ times the rms of the image represent non-associated continuum radio emission. The black circle in the bottom left corner represents the beam size of MeerKAT in each case. For sources that were observed for more than one epoch, we indicate the epoch as part of the FRB name.}
    \label{fig:non-detection}
\end{figure*}

\begin{figure*}
\centering

\begin{minipage}{1.0\textwidth}
    \centering
        \includegraphics[width=1.0\textwidth]{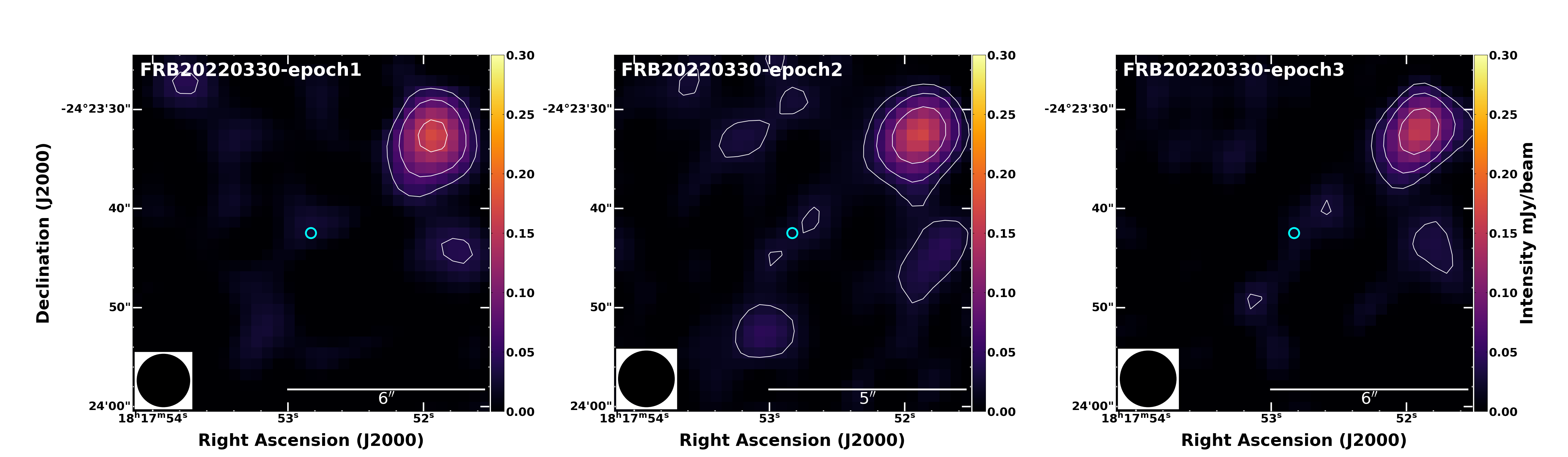}
    \end{minipage}

\begin{minipage}{1.0\textwidth}
    \centering
        \includegraphics[width=1.0\textwidth]{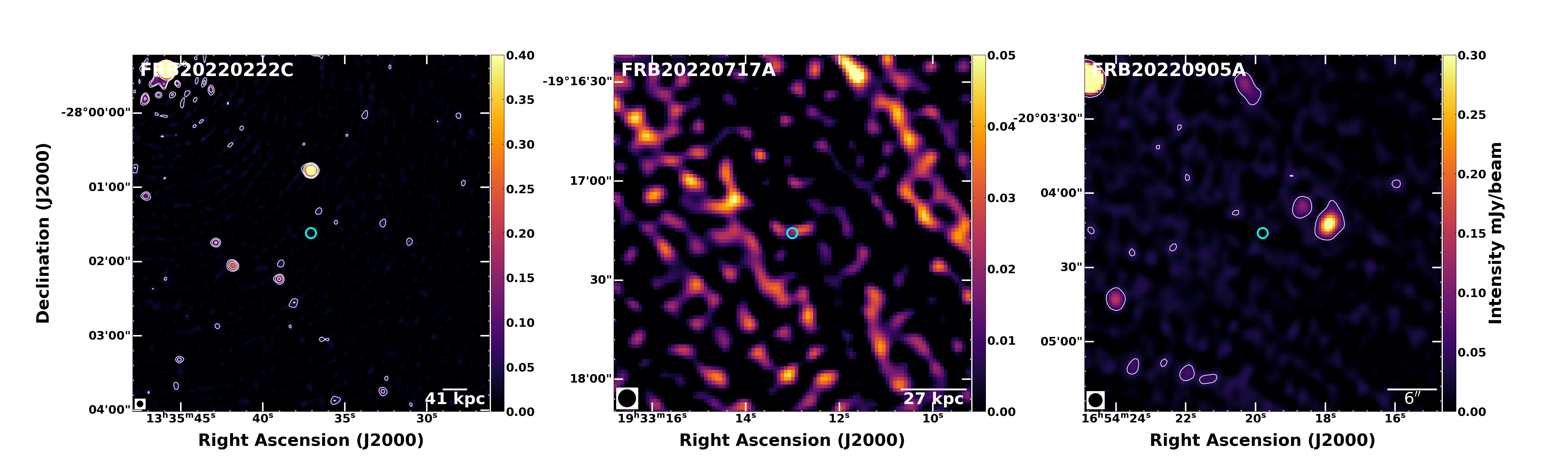}
    \end{minipage}

\begin{minipage}{1.0\textwidth}
    \centering
        \includegraphics[width=1.0\textwidth]{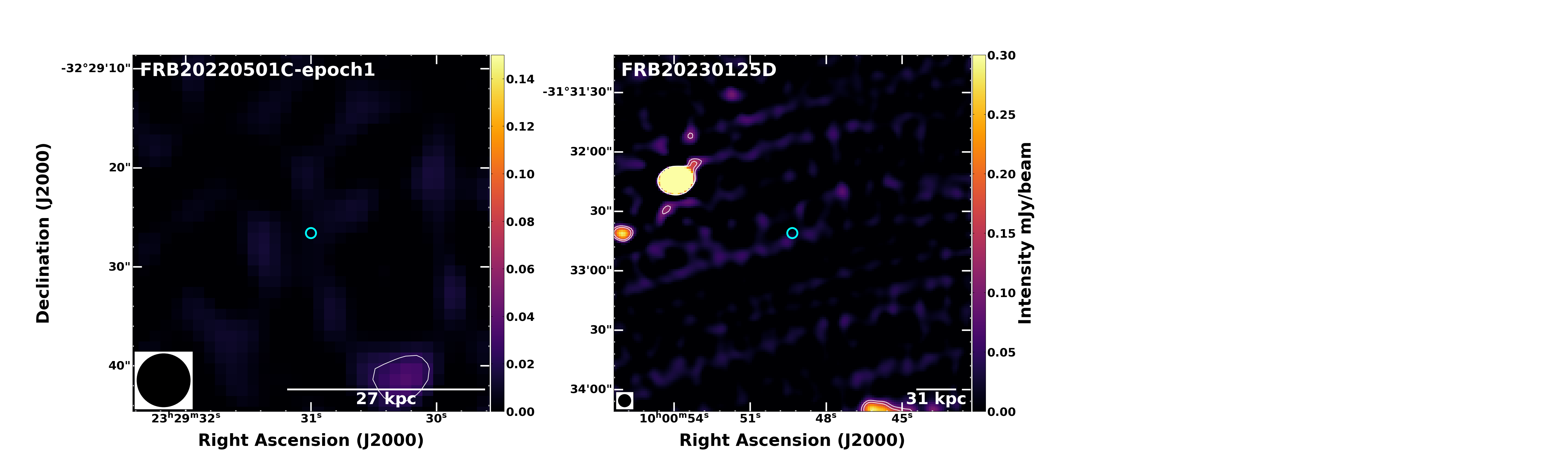}
    \end{minipage}    

    \caption{Non-detections by MeerKAT. The cyan circle indicates the position of the ASKAP/MeerTRAP FRB. White contours corresponding to $3,6,12,24$ times the rms of the image represent non-associated continuum radio emission. The black circle in the bottom left corner represents the beam size of MeerKAT in each case. For sources that were observed for more than one epoch, we indicate the epoch as part of the FRB name.}
    \label{fig:non-detection1}
\end{figure*}

%\newpage
\section{Results}\label{results}

We report on a search for PRS candidates associated with 25 FRB positions (see Appendix~\ref{tab:appendix}) using MeerKAT. These are one-off FRBs, which have been localised by ASKAP and MeerTRAP to an arcsecond position. Of the 25 FRBs, there are multi-epoch observations of some sources, which should have resulted in a total of 39 FRB fields. However, due to a calibration failure of one field, we have 38 FRB fields in total. Figures~\ref{fig:non-detection}, \ref{fig:non-detection1}, \ref{fig:detected}, and \ref{fig:detected2} show the 38 FRB MeerKAT fields we searched for PRS candidates. The ASKAP / MeerTRAP FRB position is indicated by a cyan circle, and white contours represent MeerKAT radio continuum emission. 

\subsection{MeerKAT detection of PRS candidates}
\begin{figure*}
\centering
    \begin{minipage}{1.0\textwidth}
    \centering
        \includegraphics[width=1.0\textwidth]{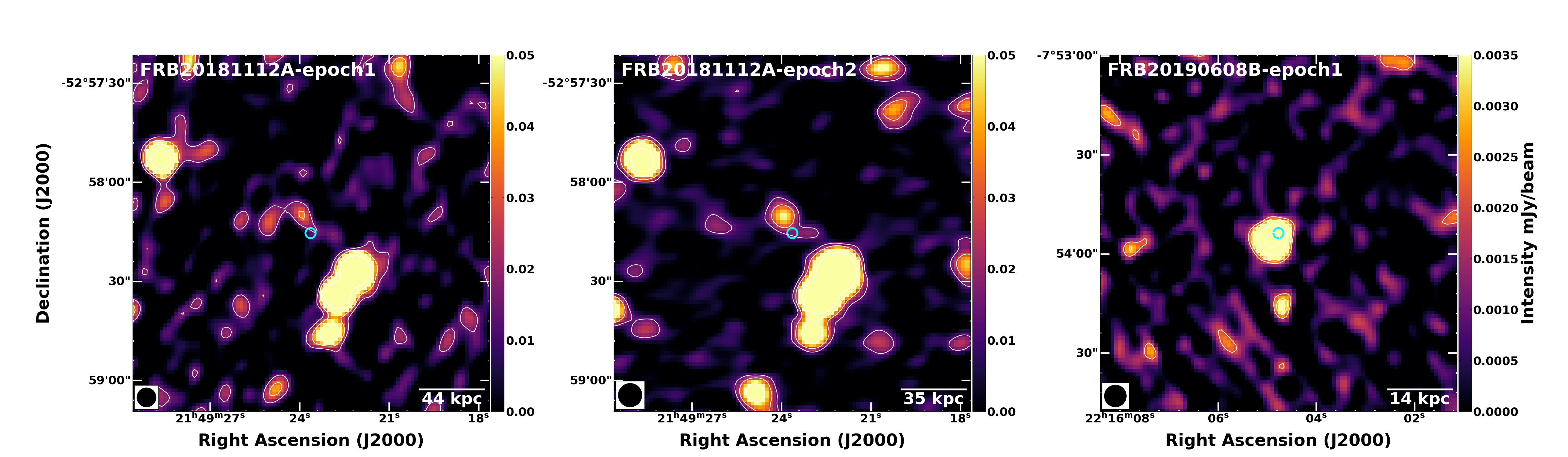}
    \end{minipage}

    \begin{minipage}{1.0\textwidth}
    \centering
        \includegraphics[width=1.0\textwidth]{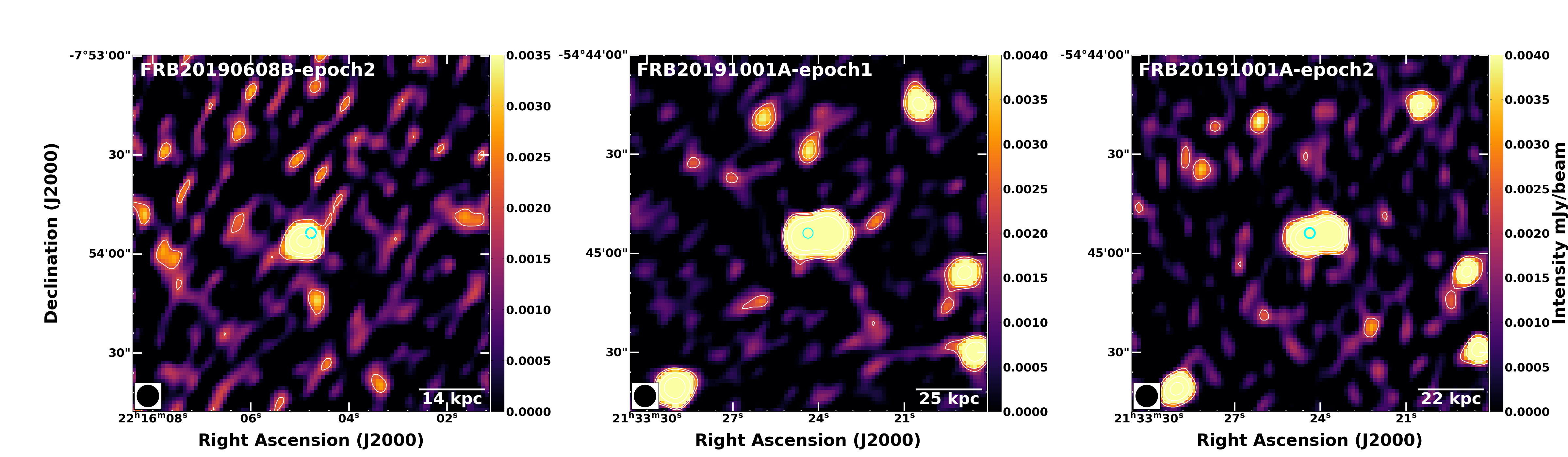}
    \end{minipage}

     \begin{minipage}{1.0\textwidth}
    \centering
        \includegraphics[width=1.0\textwidth]{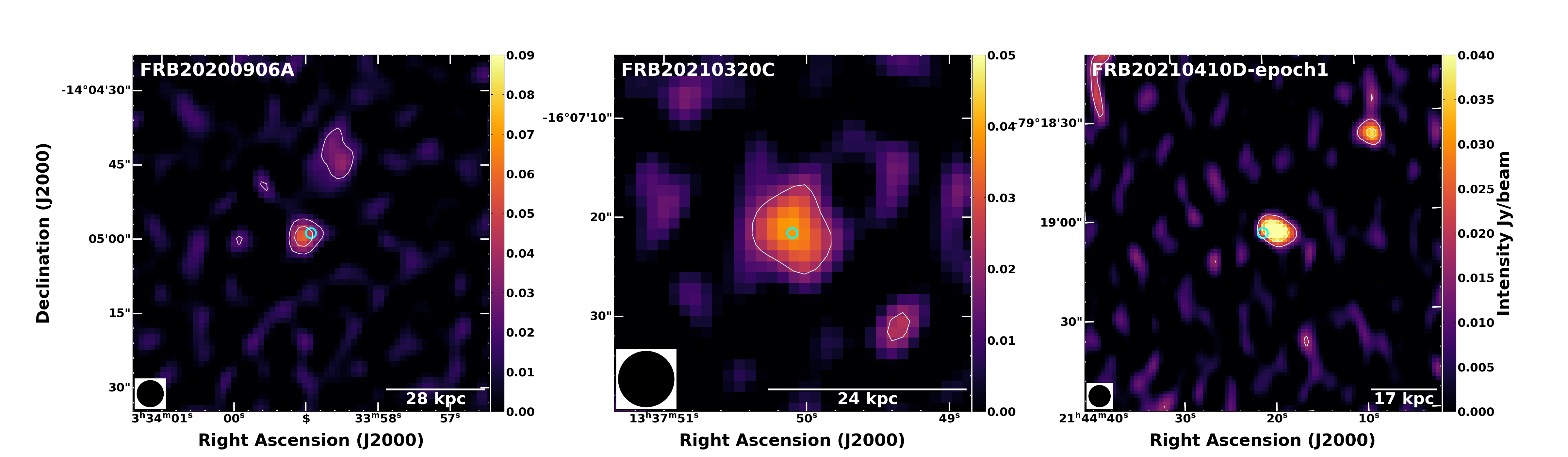}
    \end{minipage}
        
    \begin{minipage}{1.0\textwidth}
    \centering
        \includegraphics[width=1.0\textwidth]{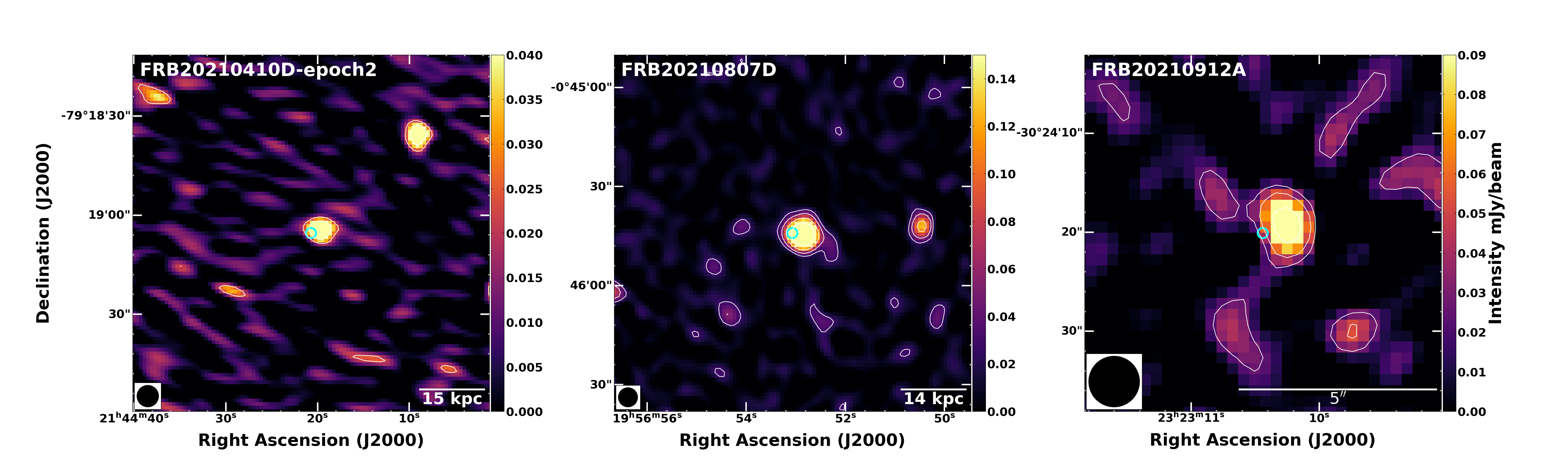}
    \end{minipage}

    \caption{MeerKAT images of various FRB sources at different angular scales. The cyan circle indicates the position of the FRB. White contours corresponding to 3, 6, 12, 24 $\times$ the respective rms (see Table \ref{tab:info1}) of the images represent continuum radio emission coincident with the FRB position. The black circle in the bottom left corner represents the beam size of MeerKAT. For sources that were observed for more than one epoch, we indicate the epoch as part of the FRB name.}
    \label{fig:detected}
\end{figure*}

\begin{figure*}
\centering
    \begin{minipage}{1.0\textwidth}
    \centering
        \includegraphics[width=1.0\textwidth]{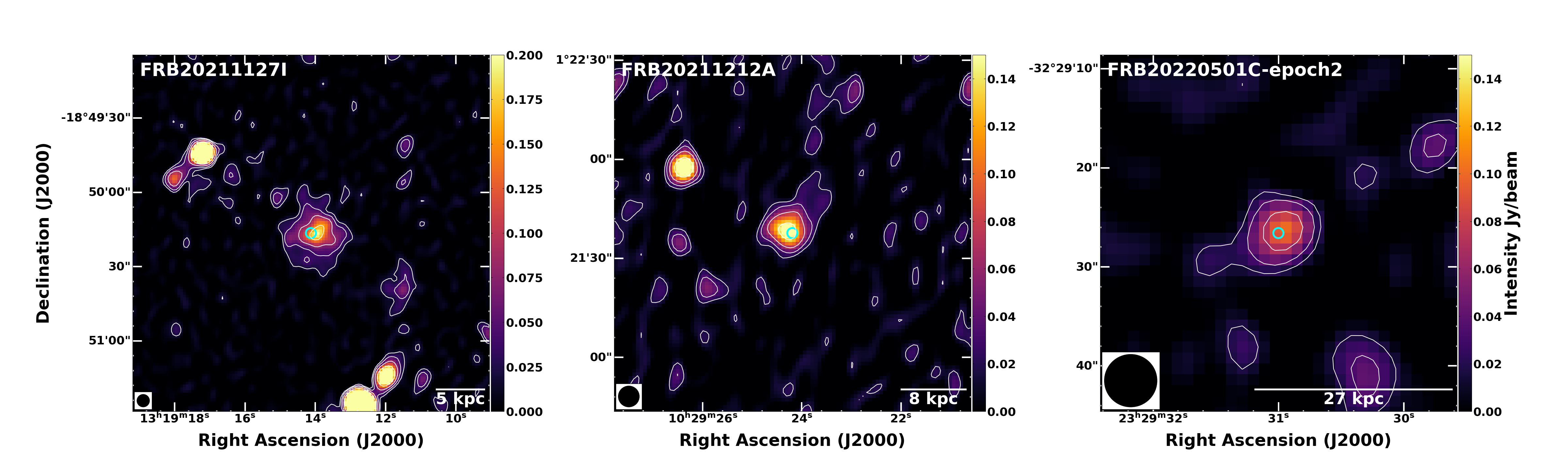}
    \end{minipage}

     \begin{minipage}{1.0\textwidth}
    \centering
        \includegraphics[width=1.0\textwidth]{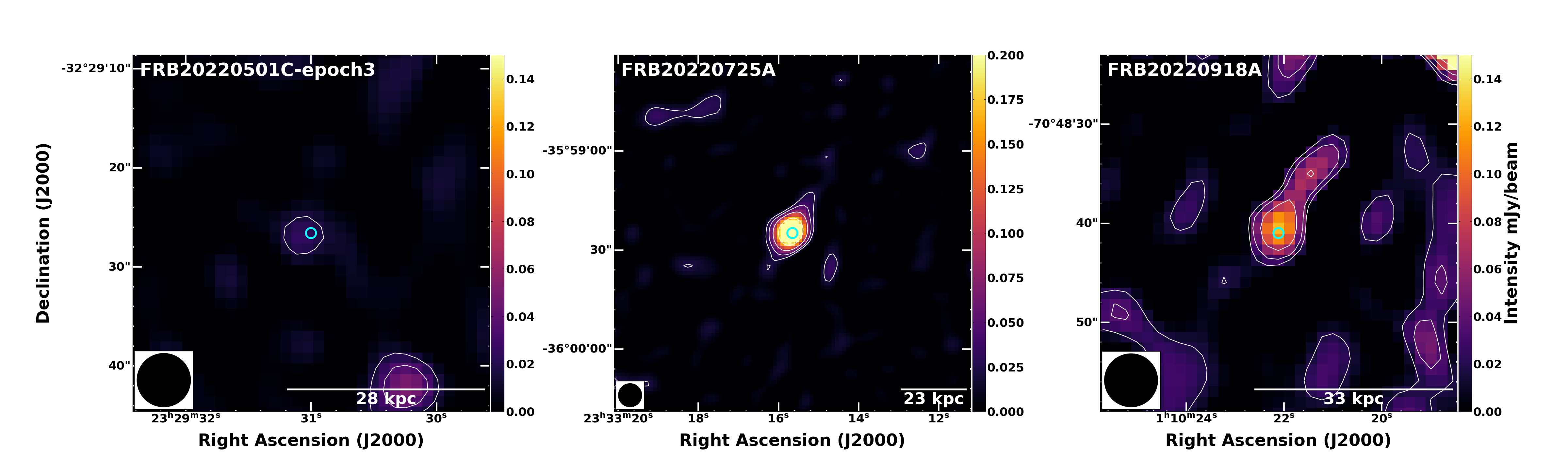}
    \end{minipage}

    \begin{minipage}{1.0\textwidth}
    \centering
        \includegraphics[width=1.0\textwidth]{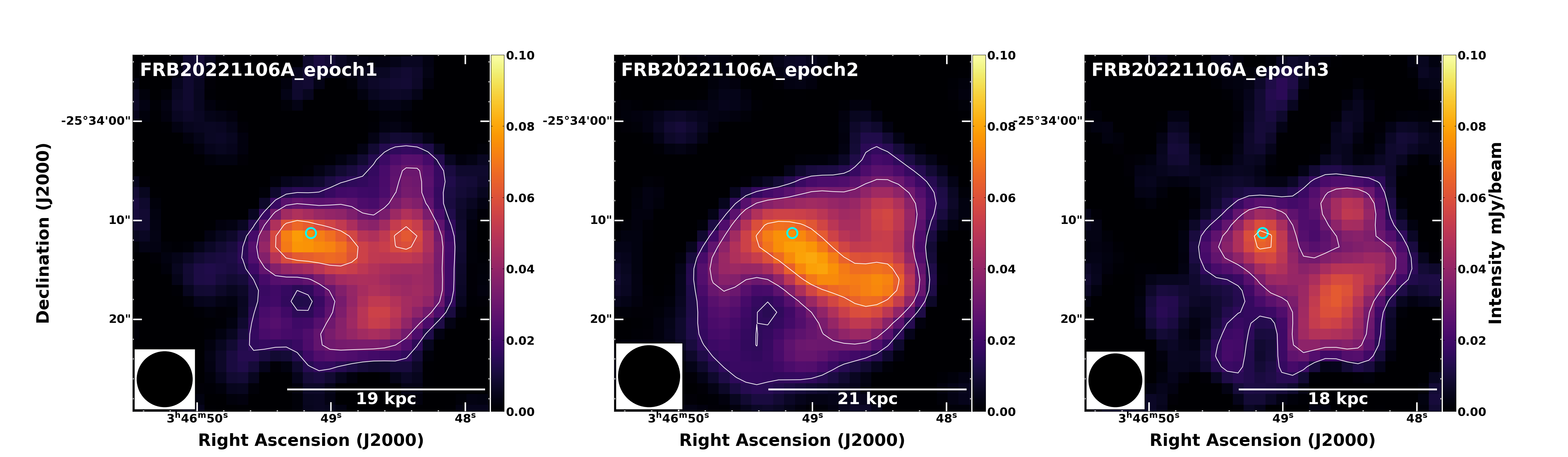}
    \end{minipage}
    
    %\begin{minipage}{1.0\textwidth}
   % \centering
    %    \includegraphics[width=1.0\textwidth]{img/detected2.png}
   % \end{minipage}

    \caption{MeerKAT images of various FRB sources at different angular scales. The cyan circle indicates the position of the FRB. White contours corresponding to 3, 6, 12, 24 $\times$ the respective rms (see Table \ref{tab:info1}) of the images represent continuum radio emission coincident with the FRB position. The black circle in the bottom left corner represents the beam size of MeerKAT. For sources that were observed for more than one epoch, we indicate the epoch as part of the FRB name.}
    \label{fig:detected2}
\end{figure*}

Out of 25 FRB positions (see Appendix~\ref{tab:appendix}), we detect radio emission coinciding with 14 FRB positions (Figures~\ref{fig:detected} and \ref{fig:detected2}), with the intensity $\geq3\sigma_{\rm rms}$ noise level. There are a total of 21 fields with detections when follow-up epochs are included. Specifically, the FRB20220501C fields, which have three follow-ups, show flux variability and a source is only detected in two observational epochs (and not detected in one epoch). The peak intensity and flux density for each detection are obtained by performing a 2D Gaussian fit of an ellipse region surrounding the detected continuum emission. Also, the flux density uncertainties were obtained and the values are reported in Table~\ref{tab:info1}. Table \ref{tab:offsets} presents the angular separations between each FRB and its associated radio source. \cite{2017Eftekhari} investigated FRB–host galaxy associations as a function of apparent magnitude and localisation region. applying their methodology to several radio facilities, including the MeerKAT telescope. Their analysis showed that MeerKAT localisation capabilities ranges to $\sim 2^{\prime\prime} -10^{\prime\prime}$. In our detection sample, 13 of the 14 radio detections are coincide with corresponding FRB positions within the expected uncertainties. One candidate which exhibits a significantly large offset, indicates poor positional coincidence between FRBs and the corresponding radio sources, despite an apparent coincidence upon visual inspection. Such a discrepancy may arise from incorrect centroiding, due to shifted source centroids caused by complex morphologies thus the position may not reflect the centroid of the relevant physical association. Consequently, we do not completely rule out the plausibility of these associations, but we regard them with caution. The physical origin of the detected radio emission remains uncertain; therefore, follow-up observations with a higher angular resolution will be essential for identifying the true nature of the emission.
 
\subsection{MeerKAT non-detection of PRS candidates}
We could not detect any radio emission coinciding with the other 11 FRB positions, and additionally also for one epoch of observation on FRB20220501C. Considering multiple epochs of observation, this should give a total 18 FRB fields. Of these, there are 17 non-detections with 1 calibration failure. Given that each FRB field has a residual image, the $\sigma_{\rm rms}$ noise of each field is obtained. We therefore compute the $3\sigma$ intensity upper limit based on the $\sigma_{\rm rms}$ and these values are reported in Table~\ref{tab:info2}. 

%  In the case of sources observed for different epochs, which are FRBs 20190102, 20220330, 20190611, 20181112 and 20200430. No continuum emission is detected near the FRBs for all epochs.

\subsection{Multi-wavelength data for potential PRSs}

\subsubsection{Optical counterparts}
%Search for optical sources consistent with the FRB position.
We searched for optical counterparts from the Panoramic Survey Telescope and Rapid Response System (Pan-STARRS) catalog (PS1) \citep{flewelling2020pan} for the $g$-band; and Dark Energy Spectroscopy Instrument (DESI), data release 10 (DR10) Legacy Imaging Survey photometric catalog \citep{dey2019overview} for the $grz$-bands. We found that optical counterparts spatially coincide with only 13 detected radio sources, as shown in Figures~\ref{fig:Optical-counterparts1} and \ref{fig:Optical-counterparts} and there is no optical source spatially coinciding with FRB20210912A. The relevant citations to detailed studies on the optical sources are listed in Table~\ref{tab:Optical}. A more detailed study towards the positions of FRB20220501, FRB20221106, FRB20220918, and FRB20210912A will be done in an upcoming work (Mfulwane et al., in prep). The Pan-STARRS and DESI catalogs are used to verify that the optical sources are not characterised as probable stars and to visually inspect the offsets between the optical and radio sources. Figures~\ref{fig:Optical-counterparts1} and \ref{fig:Optical-counterparts} also indicate the FRB position with respect to the optical source, and whether the FRB location is offset from the optical position. By visually inspecting the morphology of detected radio and optical sources, it can be seen that most the radio sources are large-scale structures, probably associated with the host galaxy. Also to confirm the coincidence between the radio and optical sources, the positional angular offsets ($\Delta \theta$) are listed in Table \ref{tab:offsets}, and some of our optical sources are resolved. We adopted the radio-optical positional offset threshold $\Delta \theta \leq 0.5^{\prime\prime}$ for the radio source to be consistent with an AGN \citep{2013Orosz,ibik2024}. For a single-component radio source the radio-optical offset threshold required ranges within $ \sim 0.5^{\prime \prime} - 1.0^{\prime\prime}$ \citep{Magliocchetti_2002, 2022Heywood}. In the case of a faint source, a threshold of $\sim 2^{\prime\prime}$ is applied, while faint and slightly extended radio sources may require the threshold of $\sim 3^{\prime\prime}$ \citep{2014Lindsay,whittam2023}. Offsets larger than $3^{\prime\prime}$ are likely due to unresolved multiple sources or complex morphology (see FRB20181112 in Figure \ref{fig:Optical-counterparts}); visual inspection shows that the radio and the optical emission coincide, thus suggesting large offsets arise from uncertainty of the centroid positions. In the case of FRB20210912, there is no optical counterpart found in the archive, thus radio-optical positional offsets are not calculated. Three of our sources are may be consistent with AGN.
 The redshift of our detected sample of radio sources range from $z=0.0469 - 0.491$, corresponding to projected sizes probed by MeerKAT that range from $\sim 5 - 40$ kpc.

\subsubsection{X-ray counterparts}
%Search for x-ray sources consistent with both the optical source and FRB position. In addition, 
We searched \textit{Chandra} \citep{weisskopf2002overview} and \textit{Swift} \citep{burrows2005swift} catalogs for X-ray counterparts to the detected radio sources. Of the 14 FRB detected sources, only one plausible X-ray counterpart was found, as shown in Figures~\ref{fig:chandra-xray} and \ref{fig:swift-xray}, subject to follow-up analyses. Confirmed X-ray counterparts would indicate the operation of non-thermal processes. Future observations or deeper searches may reveal X-ray emission for the other radio sources.

\subsubsection{Cross-check of multi-wavelength counterparts}
To ensure the robustness of the multi-wavelength associations, we performed an independent cross-check of the FRB fields within a radius of 15$''$ from each MeerKAT detection. This verification used two additional datasets: Wide-field Infrared Survey Explorer (WISE) \citep{wright2010wide} mid-infrared imaging, W1 and W2 bands, and Gaia DR3 astrometric catalogs \citep{brown2021gaia}. The procedure confirmed all optical identifications reported above and revealed a few cases where bright foreground stars are present within the search region, as identified through Gaia parallax and proper motion measurements. In particular, in the fields of FRB20220501C, FRB20181112A, and FRB20210807D, Gaia detects nearby stars located close to the radio position. While these stars are not proposed as counterparts to the FRBs, their presence highlights the possibility of contamination in the optical or infrared images, and was taken into account when visually inspecting the multi-wavelength data.

\subsection{Radio-to-optical (RO) ratio calculation}
Since most of our MeerKAT detected radio sources are not offset from their likely optical counterparts (hosts) and seem to be consistent with large-scale structures, we used the RO ratio as a diagnostic metric to check for ``radio excess'' above the expected emission due to star formation, as given in Table~\ref{tab:RO-ratio}. A significant excess radio emission could signal the presence of non-thermal radio emission (a PRS). The RO ratio is calculated as $RO=\log_{10}(S_{\rm 1.2 GHz}/S_{\rm optical})$, where $S_{\rm 1.2 GHz}$ is the flux density of MeerKAT radio sources at 1.2~GHz in Jansky and $S_{\rm optical}$ is the optical flux density in Jansky. The optical flux density is calculated as $S_{\rm optical} = 3631 \times 10^{-0.4m_{\rm r}}$ \citep{2005Afonso, ibik2024}, where $m_{\rm r}$ is the published AB $r$-band magnitude of the host galaxies. We adopted the threshold used by \citet{Eftekhari_2021,ibik2024}, where an RO ratio $< 1.4$ is consistent with pure star formation and an RO ratio $> 1.4$ is consistent with other radio emission such as those related to AGN activities, pulsar wind nebulae, hypernebulae, supernova remnants, gamma-ray bursts, etc. For example, \citet{ibik2024} found an RO ratio of the PRS associated with FRB20121102A of $\sim 2.9$ using 250 $\mu$Jy at 1.63 GHz and $\sim 1.7$ using 258 $\mu$Jy at 1.5 GHz for the PRS associated with FRB20190520B \citep{ibik2024}, which imply radio emissions that are not solely due to star formation but could be linked to AGN or PRS emission. Of the 14 detected sources, we performed the calculation for only nine sources for which we could find the published AB-magnitudes. For the sources for which we could perform the calculation, the ratio is $< 1.4$ suggesting that the radio emission is consistent with pure star-formation (see Table~\ref{tab:RO-ratio}).

\begin{table*}
\centering
 \caption{Detection of continuum radio emission.}
 \label{tab:info1}
 \rotatebox{90}{
 \begin{tabular}{cccccccccc}
  \hline
  Source name & Observation date & R.A.(J2000) & Dec.(J2000) & Synthesised beam & rms  & Peak Flux &Maj$\times$Min axis&Pos. Angle&Int. Flux\\
  & & & & & (mJy beam$^{-1}$) & mJy beam$^{-1}$ & & & mJy\\
    \hline 
    FRB20181112A & 19-Apr-2021 &  21:49:23.63 &-52:58:15.4&7\arcsec.304$\times$ 7\arcsec.304&0.0047& 0.0371& 5\arcsec.63 $\times$ 5\arcsec.15 &45$^\circ$ &$0.0202\pm0.0091$\\
  FRB20181112A & 03-Sep-2021 &  21:49:23.63 &-52:58:15.4&5\arcsec.940$\times$ 5\arcsec.940&0.0054& 0.0219& 5\arcsec.64 $\times$ 3\arcsec.55 &34$^\circ$ &$0.0124\pm 0.0061$\\
  \hline
  FRB20190608B & 10-Apr-2021 & 22:16:04.77 &-07:53:53.7 &6\arcsec.621$\times$ 6\arcsec.621&0.000581&0.0170&7\arcsec.45 $\times$ 6\arcsec.87&146$^\circ$&$0.0199\pm 0.0014$\\
  FRB20190608B & 02-Sep-2021 & 22:16:04.77 & -07:53:53.7& 6$\arcsec$.465$\times$ 6\arcsec.465&0.000587& 0.0174& 7\arcsec.36 $\times$ 6\arcsec.78& 82$^\circ$ & $0.0205 \pm 0.0015$\\
  \hline
  FRB20191001A &  10-Apr-2021 & 21:33:24.41&  -54:44:53.9 &6\arcsec.831$\times$ 6\arcsec.831&0.00057&0.0571&11\arcsec.97$\times$6\arcsec.91&98$^\circ$&$0.1013 \pm 0.0110$\\
  FRB20191001A & 02-Sep-2021 & 21:33:24.41& -54:44:53.9& 5\arcsec.914$\times$ 5\arcsec.914& 0.00067& 0.0462 & 11\arcsec.25$\times$5\arcsec.63 & 98$^\circ$ & $0.0837 \pm 0.0130$\\
  \hline
  FRB20200906A &  18-Jan-2023 & 03:33:58.93 &-14:04:58.8 &5\arcsec.476$\times$ 5\arcsec.476&0.007175&0.0521&5\arcsec.43$\times$ 5\arcsec.00&134$^\circ$&$0.0473 \pm 0.0069$\\
  \hline
  FRB20210320C & 19-Feb-2023 &13:37:50.10  & -16:07:21.6 &5\arcsec.698$\times$ 5\arcsec.698,&0.006837&0.0317&7.\arcsec.49$\times$6\arcsec.20&27$^\circ$&$0.0454\pm 0.0069$\\
  \hline
  FRB20210410D &  05-Sep-2021 & 21:44:20.7 & -79:19:05.5&6\arcsec.723$\times$ 6\arcsec.723&0.0053&0.0546&8\arcsec.22 $\times$ 5\arcsec.87&63$^\circ$&$0.0583\pm 0.0080$\\
  FRB20210410D & 14-Feb-2023 & 21:44:20.7 &  -79:19:05.5 & 5\arcsec.880$\times$ 5\arcsec.880& 0.0076&0.0714 & 6\arcsec.09 $\times$ 5\arcsec.24 & 98$^\circ$&$0.0659\pm 0.0074$\\
  \hline
  FRB20210807D & 23-Dec-2022 & 19:56:53.07 & -00:45:44.1 &5\arcsec.997$\times$ 5\arcsec.997&0.0089&0.4699&6\arcsec.40$\times$6\arcsec.11&60$^\circ$&$0.5110\pm 0.0260$\\
  \hline
  FRB20210912A & 22-Dec-2022 &23:23:10.44  &-30:24:20.1  &5\arcsec.171$\times$ 5\arcsec.171&0.0084& 0.1318& 5\arcsec.16$\times$3\arcsec.76 &18$^\circ$&$0.0957 \pm 0.0106$\\
  \hline
  FRB20211127I & 18-Jan-2023 & 13:19:14.12  & -18:50:16.5  &5\arcsec.299 $\times$ 5\arcsec.299&0.0066&0.1495&9\arcsec.78 $\times$ 7\arcsec.91&120$^\circ$&$0.4118\pm 0.0800$\\
  \hline
   FRB20211212A & 15-Dec-2022 & 10:29:24.19 & +01:21:37.6  &6\arcsec.526$\times$ 6\arcsec.526&0.0058&0.1596&8\arcsec.26$\times$7\arcsec.79&63$^\circ$&$0.2413 \pm 0.0360$\\
  \hline
   FRB20220501C & 25-Nov-2023 & 23:29:31.00 &-32:29:26.60&5\arcsec.388$\times$ 5\arcsec.388&0.0056&0.0899 &4\arcsec.71$\times$ 4\arcsec.03&136$^\circ$&$0.0588 \pm 0.0130$\\
  FRB20220501C & 26-Nov-2023 & 23:29:31.00 &-32:29:26.6&5\arcsec.480$\times$ 5\arcsec.480&0.0053&0.0609&5\arcsec.95$\times$5\arcsec.72&98$^\circ$&$0.0691\pm 0.0038$\\
  \hline
  FRB20220725A & 09-Nov-2023 &23:33:15.65 &-35:59:24.9 &7\arcsec.268$\times$ 7\arcsec.268&0.0072&0.3506& 8\arcsec.11$\times$6\arcsec.45 & 147$^\circ$ & $0.3473\pm 0.0370$\\
  \hline 
  FRB20220918A &  16-Nov-2023 & 01:10:22.11 &-70:48:41.0 &5\arcsec.433$\times$ 5\arcsec.433&0.0058&0.0859&3\arcsec.36$\times$2\arcsec.21&163$^\circ$&$0.0216\pm 0.0025$\\
  \hline
   FRB20221106A & 21-Nov-2023 & 03:46:49.15&-25:34:11.3&5\arcsec.649$\times$ 5\arcsec.649&0.0048&0.0446&14\arcsec.30$\times$5\arcsec.33&78$^\circ$&$0.1065\pm 0.0480$\\
  FRB20221106A & 05-Dec-2023 & 03:46:49.15&-25:34:11.3&6\arcsec.264$\times$ 6\arcsec.264&0.0054&0.0557&18\arcsec.22$\times$8\arcsec.37&67$^\circ$&$0.2167\pm 0.0670$\\
  FRB20221106A & 20-Jan-2024 & 03:46:49.15&-25:34:11.3&5\arcsec.452$\times$ 5\arcsec.452&0.0053&0.0270&20\arcsec.59$\times$5\arcsec.39&55$^\circ$&$0.1096\pm 0.0700$\\
  \hline
  \end{tabular}}
    \label{tab:my_label}
\end{table*}

\begin{table*}
\centering
 \caption{Sources for which continuum radio emission is not detected.}
 \label{tab:info2}
 \begin{tabular}{ccccccc}
  \hline
  Source name & Observation & R.A.(J2000) & Dec.(J2000) & Synthesised beam & rms  & Upper limit\\
   & date & &  &  & mJy beam$^{-1}$ & mJy beam$^{-1}$\\
  \hline
   FRB20190102C & 10-Apr-2021 &21:29:39.76 &-79:28:32.5&5\arcsec.709$\times$ 5\arcsec.709&0.0054&$< 0.0162$\\
  FRB20190102C &05-Sep-2021 &21:29:39.76 &-79:28:32.5&6\arcsec.723$\times$ 6\arcsec.723&0.0055&$< 0.0164$\\
  \hline
  FRB20190611B & 10-Apr-2021 & 21:22:58.94 &-79:23:51.3&6\arcsec.446$\times$ 6\arcsec.446&0.0052&$< 0.0155$\\
  FRB20190611B & 05-Sep-2021 & 21:22:58.94 &-79:23:51.3&6\arcsec.723$\times$ 6\arcsec.723&0.0055&$< 0.0164$\\
  \hline
  FRB20191228A & 06-Sep-2021 & 22:57:43.33  & -29:35:38.8 & 5\arcsec.813$\times$ 5\arcsec.813& 0.0063& $<0.0189$\\
  \hline
  FRB20220330 &  20-Jan-2023 & 18:17:52.83 &-24:23:42.5&5\arcsec.746$\times$ 5\arcsec.746&0.0120&$< 0.0360$\\
   FRB20220330 &  01-Dec-2023 & 18:17:52.83 &-24:23:42.5&5\arcsec.413$\times$ 5\arcsec.413&0.0080&$< 0.0241$\\
  FRB20220330 &  13-Jan-2024 & 18:17:52.83 &-24:23:42.5&5\arcsec.728$\times$ 5\arcsec.728&0.0089&$<0.0269 $\\
   \hline
   FRB20200430A & 19-Apr-2021 & 15:18:49.55 &+12:22:34.8&7\arcsec.283$\times$ 7\arcsec.283& 0.0058& $<0.0175$\\
  FRB20200430A & 28-Jul-2021 &  15:18:49.55 &+12:22:34.8&8\arcsec.583$\times$ 8\arcsec.583& 0.0056& $< 0.0164$\\
  \hline
  FRB20210117A & 16-Dec-2022 &22:39:55.015 & -16:09:05.45&5\arcsec.185 $\times$ 5\arcsec.185& 0.0068&$< 0.0205$\\
  \hline
  FRB20220105A & 08-Jan-2023 &13:55:12.81  & +22:27:58.4  &6\arcsec.350$\times$ 6\arcsec.350&0.0065&$<0.0196$ \\
  \hline
  FRB20220222C &   09-Feb-2023 & 13:35:37.06 &-28:01:37.2&5\arcsec.382$\times$ 5\arcsec.382&0.0059& $<0.0179$\\
  \hline
   FRB20220501C & 09-Nov-2023 & 23:29:31.00 &-32:29:26.6&5\arcsec.441$\times$ 5\arcsec.441&0.0056&$< 0.0168$\\
   \hline
   FRB20220717A &  28-Dec-2023 & 19:33:13.0 & -19:17:15.8 &--&--&Calibration failed\\
   FRB20220717A &  02-Mar-2024 & 19:33:13.0 & -19:17:15.8 & 5\arcsec.419$\times$ 5\arcsec.419& 0.0095 & $< 0.0284$\\
  \hline
  FRB20220905A & 12-Dec-2023 & 16:54:19.79 & -20:04:16.2 &5\arcsec.644$\times$ 5\arcsec.644&0.0017&$< 0.0052$\\
  \hline
  FRB20230125D & 18-Jun-2024 & 10:00:49.33 &-31:32:41.1&6\arcsec.606 $\times$ 6\arcsec.606&0.0059&$< 0.0179$\\
  \hline
    \end{tabular}
    \label{tab:non-detection}
\end{table*}

\begin{table*}
\centering
 \caption{List of redshifts of the FRB as well as the optical counterpart.}
 \label{tab:Optical}
 \begin{tabular}{cccccccc}
  \hline
  Source name & Optical  & Ref. & X-ray & Host(As reported)\\
              &Counterpart &(Host) &  counterpart & \\
  \hline
  FRB20181112A & Yes(DESI-Legacy-survey)& \cite{Prochaska2019}&No& Host galaxy\\
  \hline
  FRB20190608B  &Yes (Pan-STARRS)& \cite{Bhandari_2022}&Yes (Chandra) & AGN(seyferts)\\
  \hline
  FRB20191001A & Yes (DESI-Legacy-survey )& \cite{Bhandari_2020a}&No&Radio galaxy \\
  \hline
  FRB20200906A & Yes (Pan-STARRS)& \cite{Bhandari_2022}&No&SF galaxy\\
  \hline
  FRB20210320C &Yes (Pan-STARRS)& \cite{gordon2023}&No&SF galaxy\\
  \hline
  FRB20210410D & Yes (DESI-Legacy-survey)&\cite{Caleb-2023}&No&Quiescent galaxy\\
  \hline
  FRB20210807D &Yes(Pan-STARRS)& \cite{gordon2023}&No&Quiescent galaxy \\
  \hline
  FRB20210912A & No &\cite{Marnoch2023}&No & Unseen host galaxy \\
  \hline
   FRB20211127I & Yes (Pan-STARRS)& \cite{gordon2023}&No&SF galaxy\\
  \hline
   FRB20211212A  & Yes (Pan-STARRS)&\cite{gordon2023}&No&Host galaxy\\
  \hline
  FRB20220501C & Yes (DESI-Legacy-survey) & \cite{Shannon2025}&No& Host galaxy\\
  \hline
  FRB20220725A &Yes(DESI-Legacy-survey)&\cite{Shannon2025}&No&host galaxy\\
  \hline
  FRB20220918A &Yes (DESI-Legacy-survey)& \cite{Shannon2025}& No & LMC\\
  \hline
  FRB20221106A &Yes (Pan-STARRS)& \cite{Shannon2025}&No& Host galaxy\\
  \hline
  \end{tabular}
    \label{tab:optical}
\end{table*}

\begin{table*}
\centering
 \begin{tabular}{ccccccc}
  \hline
  Source name & \multicolumn{2}{c|}{Position in radio}   &Radio-FRB offsets  &  \multicolumn{3}{c|}{Radio-Optical offsets} \\
              &  RA   &   Dec  & $\Delta \theta_{FRB}$ \big($^{\prime\prime}$\big) & $\Delta \theta_{Radio} $\big($^{\prime\prime}$\big)& $\Delta$RA \big($^{\prime\prime}$\big)& $\Delta$Dec \big($^{\prime\prime}$\big)\\
    \hline
    FRB20181112A &21:49:24.000  
 & -52:58:09.859 & 6.5& 6.6& 5.51 & -5.69\\
  \hline
  FRB20190608B &22:16:04.882 
 &-7:53:55.556& 2.49& 0.18&-0.029&0.177\\
  \hline
  FRB20191001A &21:33:24.527 
  &-54:44:54.598& 1.23& 0.59& 0.553 & 0.212\\
  \hline
  FRB20200906A & 3:33:59.033 
 &-14:04:59.107& 1.53& 0.78& 0.77 & 0.172\\
  \hline
  FRB20210320C &13:37:50.014 
 &-16:07:22.510&1.54 & 1.9&1.684&0.837\\
  \hline
   FRB20210410D &21:44:19.969  
 &-79:19:04.868&2.15 & 3.0 & 3.00 & 0.045 \\
  \hline
  FRB20210912A & 23:23:10.243 
&-30:24:19.732 &2.57 & ----& ----&----\\
  \hline
  FRB20210807D &  19:56:52.920 
 &-0:45:44.500&2.29 & 0.19 & -0.075 & -0.172\\
  \hline
   FRB20211127I & 13:19:13.848  
 &-18:50:18.900&  3.96& 1.65& 1.62& 0.34\\
  \hline
   FRB20211212A &10:29:24.290 
 &1:21:38.800& 1.92& 1.19&1.065 &-0.525 \\
  \hline
   FRB20220501C &23:29:30.999  
&-32:29:26.584& 0.020& 0.43& 0.404&0.154\\
  \hline
   FRB20220725A &23:33:15.631 
&-35:59:24.900& 0.23& 0.62&0.595&-0.174\\
  \hline
  FRB20220918A &1:10:22.186  
&-70:48:40.505& 0.618& 1.21&1.11&0.474\\
  \hline
  FRB20221106A &3:46:48.358  
&-25:34:19.300&13.37&12.6&8.059&9.69\\
  \hline
  \end{tabular}%}
  \caption{Summary of positional offsets. The table presents (i) the angular separation $\Delta \theta_{FRB}$ between the radio sources and the FRB positions in arcseconds, and (ii) the offsets between the radio and optical source positions. The latter includes the angular separation $\Delta \theta_{Radio}$ and the $\Delta$RA and $\Delta$Dec offset components in arcseconds.}
    \label{tab:offsets}
\end{table*}

\begin{table*}
\centering
 \begin{tabular}{ccccccc}
  \hline
  Source name & Instrument & $m_r$ & Ref. & Optical-flux & Radio-flux  & RO-ratio \\
              &           &  mag[AB]  & &    mJy       &  mJy        & \\
    \hline 
    FRB20181112A &DES(DECam) &$21.68$& 1&0.0077 & 0.0662 & 0.93\\
  \hline
  FRB20190608B & DECaLs(DECam)&$17.41$& 2,3&0.3945 & 0.0236 & -1.22\\
  \hline
  FRB20191001A & DECaLS(DECam) &$18.36$& 4&0.1644 & 0.1059 & -0.19\\
  \hline
  FRB20200906A & DECaLS(DECam) &$19.95$& 5&0.0380 & 0.1298 & 0.53\\
  \hline
  FRB20210320C & SOAR(Goodman)&$19.47$& 3,6&0.0592 & 0.1260 &0.33\\
  \hline
   FRB20210410D & SOAR(Goodman)&$20.65$& 3,7& 0.0200 & 0.0977 & 0.69\\
  \hline
  FRB20210912A &----&----&----&----&----&----\\
  \hline
  FRB20210807D & Pan-STARRS &$17.17$& 3,6&0.4921 & 0.4970 &0.00\\
  \hline
   FRB20211127I & SOAR(Goodman) &$14.96$& 3,6 &3.7673 &0.5055&-0.87\\
  \hline
   FRB20211212A &SOAR(Goodman) &$16.44$&3,6&0.9639& 0.3600& -0.43\\
  \hline
   FRB20220501C &----&----&----&----&----&----\\
  \hline
   FRB20220725A &----&----&----&----&----&----\\
  \hline
  FRB20220918A &----&----&----&----&-----&----\\
  \hline
  FRB20221106A &----&----&----&----&----&----\\
  \hline
  \end{tabular}%}
  \caption{The optical photometry of FRB host galaxies detected by ASKAP/CRAFT from different imaging instruments. The AB-m$_r$ magnitude values in the $r-$band were obtained from the literature. References:  (1)~\citet{Prochaska2019}, (2)~\citet{Macquart2020}, (3)~\citet{gordon2023}, (4)~\citet{Bhandari_2020a}, (5)~\citet{Bhandari_2020b}, (6)~\citet{james2022}, (7)~\citet{Caleb-2023}.} 
    \label{tab:RO-ratio}
\end{table*}

\section{Discussion}\label{discussion}
Of the 38 FRB target fields reported in this paper, we have detected continuum radio emission toward 14 unique FRB positions (a total of 21 FRB fields for multiple observational epochs). There is no detection toward 11 FRB unique positions (a total of 18 FRB fields for multiple epochs). These non-detections might suggest that the radio sources have faded away (quiescent source), or that they are too faint for MeerKAT to detect. Therefore, a radio flux upper limit for each field was derived.  

For all 14 detected radio sources, we searched for optical counterparts in archival catalogues and found 13; of these, 11 optical counterparts have previously been studied in detail (Table~\ref{tab:optical}). Given the resolution of MeerKAT telescope, we detected unresolved radio emission and by visual inspection, our radio sources overlap with these optical sources. In addition, the morphology of the detected radio sources seems to resemble that of the corresponding optical sources, e.g., the radio emission structure of FRB20191001A is symmetric with respect to the two optical galaxies. Therefore, the detected radio sources are consistent with the large-scale optical sources instead of compact PRSs. Therefore, we applied a diagnostic method to check if the continuum radio emission is consistent with star formation or if there is an excess radio emission (see Table~\ref{tab:RO-ratio}) resulting from sources such as AGN or PRSs. For the nine sources for which we could do the calculations, the detections are consistent with star formation ($RO < 1.4$). These results may imply that the continuum emission from star formation might be dominating over any faint PRS signal that might exist. In addition, the luminosity of the MeerKAT radio sources range from $2.7 \times 10^{28} - 9.4 \times 10^{29}$ erg s$^{-1}$ Hz$^{-1}$, to check if the luminosity of these sources are consistent with the PRSs detected. Thus, the luminosities seem to fall within the range, therefore high resolution observations are required to confirm the nature of these sources. A deep targeted search with a high-resolution radio telescope is necessary to elucidate the existence of compact PRSs, resolving them from the radio emission detected here (Mfulwane et al., in prep). The lack of offset between optical and radio means we cannot exclude the possibility of an AGN origin. \footnote{Our logic is as follows. The radio emission may trace the AGN activity, while the optical emission traces the host galaxy (star formation) emission. When there is a large radio–optical offset $> 0.5^{\prime\prime}$, the radio emission  could be due to other processes such as star formation. On the other hand, a small radio-optical offset, can imply that the radio emission originates from the same region as the optical nucleus, suggesting an AGN origin. Thus, the radio emission does not originate from star formation but from AGN-driven processess such as synchrotron emission from jets or a compact core} Of all the radio sources detected, only the radio source associated with FRB20220501C shows variability behaviour in the radio band. The optical counterpart associated with this source is not classified. No comparative analysis was performed for the X-ray data, as no X-ray counterparts were detected at the relevant positions, except for a single case where it is uncertain if is a real detection.

\section{Conclusions and future work}\label{conc}
We conducted a search for continuum radio emission using MeerKAT towards 25 FRBs localised by ASKAP and MeerTRAP. This study is a precusor for searching and identifying potential PRS candidates. Of the 25 FRB positions, we detected radio sources for 14 unique FRB positions and the flux densities of these detected sources have been determined. In the case of FRB20220501C field that was observed for three epochs, we found no detection in the first epoch, but made a detection in the second and third epochs, suggesting flux variability on day-timescales. This variability is intrinsic or due to scintillation effects. We furthermore found potential optical counterparts for 13 radio sources, and a possible X-ray counterpart for only one source.
Most of the optical counterparts found in the archival catalogues are classified as galaxies (FRB host galaxies). There are detailed studies in the literature for most of these optical sources, except for optical sources coinciding with the radio emission in three FRB fields, namely, FRB20220501C, FRB20221106A, and FRB20220918A. Therefore, no identification can be made regarding these three radio sources. One may also study the unknown hosts of these three FRBs at sub-millimeter wavelengths.

%Our detected radio sources are unresolved and consistent with optical sources that have been identified as the host galaxies of FRBs 
Our detected radio sources are unresolved, overlap with the optical sources, and furthermore their morphology have some resemblance to the optical ones. This may imply that we observed the large-scale (host galaxy) emission from star formation instead of PRSs. Indeed, we found that the radio emission is consistent with star formation within the host using the RO ratio, and there is no sign of excess emission (see Table~\ref{tab:RO-ratio}) resulting from sources such as PRSs. We therefore conclude that either there is no PRS, or a faint PRS may be embedded within the radio emission resulting from star formation in the region. In order to confirm the existence of a compact PRS, deep, targeted observations using the high-resolution enhanced Multi-Element Remote-Linked Interferometer (e-MERLIN) telescope are required with the aim of disentangling the observed radio sources from compact PRSs and, if detected, determine the size of the compact PRSs, check for variability and spectral shape, constrain the flux density, and search for very high energy counterparts with the High Energy Stereoscopic System (H.E.S.S.). %We will lastly compare the flux density to that of MeerKAT. 

\section*{Acknowledgements}
This paper utilises MeerKAT L-band observational data from 2021, 2022, and 2023 open-time proposals (SCI-20210212-CV-01, SCI-20220822-CV-01, SCI-20230907-CV-01, respectively). The South African Radio Astronomy Observatory, a facility of the National Research Foundation, operates the MeerKAT telescope. The Inter-University Institute for Data Intensive Astronomy (IDIA) visualisation lab https://idia.ac.za/citing-idia-ilifu-in-publications/ is used to for this work. IDIA is a partnership of the University of Cape Town, the University of Pretoria, the University of the Western Cape and the South African Radio Astronomy Observatory. In addition we acknowledge the ilifu cloud computing facility- www.ilifu.ac.za.

%%%%%%%%%%%%%%%%%%%%%%%%%%%%%%%%%%%%%%%%%%%%%%%%%%
\section*{Data Availability}
The data used in this work will be shared upon request. 
%The inclusion of a Data Availability Statement is a requirement for articles published in MNRAS. Data Availability Statements provide a standardised format for readers to understand the availability of data underlying the research results described in the article. The statement may refer to original data generated in the course of the study or to third-party data analysed in the article. The statement should describe and provide means of access, where possible, by linking to the data or providing the required accession numbers for the relevant databases or DOIs.



\bibliographystyle{mnras}
\bibliography{example} % if your bibtex file is called example.bib

%\begin{thebibliography}{99}
%\end{thebibliography}

\newpage

%%%%%%%%%%%%%%%%%%%%%%%%%%%%%%%%%%%%%%%%%%%%%%%%%%

%%%%%%%%%%%%%%%%% APPENDICES %%%%%%%%%%%%%%%%%%%%%

\appendix \label{appendix}

\section{FRB properties}

Table \ref{tab:appendix} lists all the FRB names and their key properties discovered by ASKAP/CRAFT and MeerTRAP.

\subsection{A brief overview of individual MeerKAT detections}
This section gives a brief overview of 14 radio sources detected by MeerKAT. We provide details of some properties of the emission and their hosts. 

\subsubsection{FRB20181112A field}
This is a two-epoch observation. \citet{Prochaska2019} studied the host galaxy of the FRB in detail and identified a bright foreground galaxy close to the host galaxy. The foreground galaxy consists of a massive halo of gas, which acts as a reservoir of material that may fuel star formation in the future. Both the foreground and the host galaxies have been identified in optical wavelengths in the DESI-legacy-survey, where the northern optical source is the foreground, while the southern source is the host galaxy. Therefore, the detection of continuum radio emission is probably stronger for the foreground galaxy compared to the host galaxy. No X-ray source was found.

\subsubsection{FRB20190608B field}
\begin{figure*}
    \centering
    \begin{minipage}{0.9\textwidth}
    \centering
    \includegraphics[width=\linewidth]{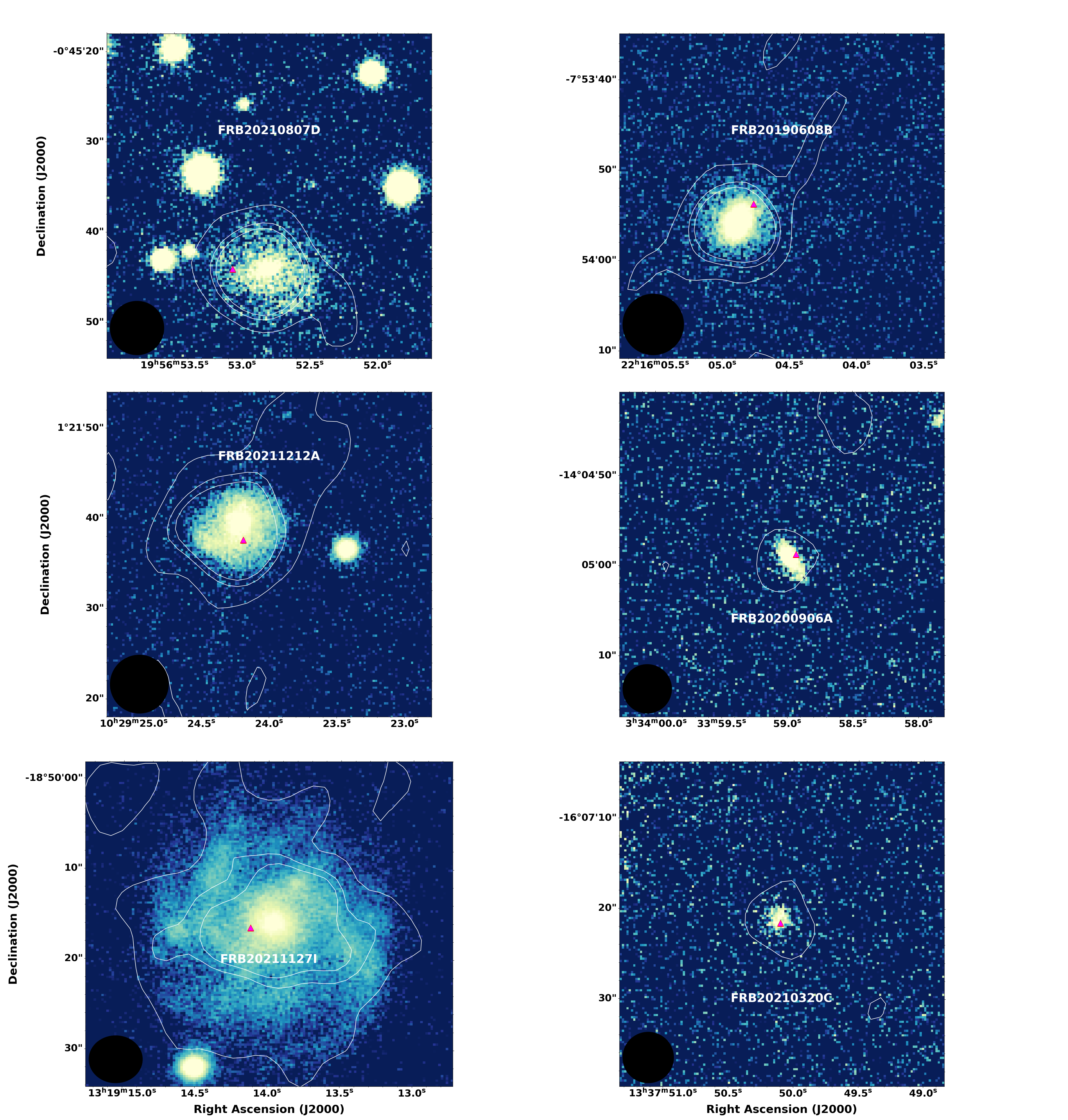}
    \end{minipage}

    \begin{minipage}{0.9\textwidth}
    \centering
    \includegraphics[scale=0.3]{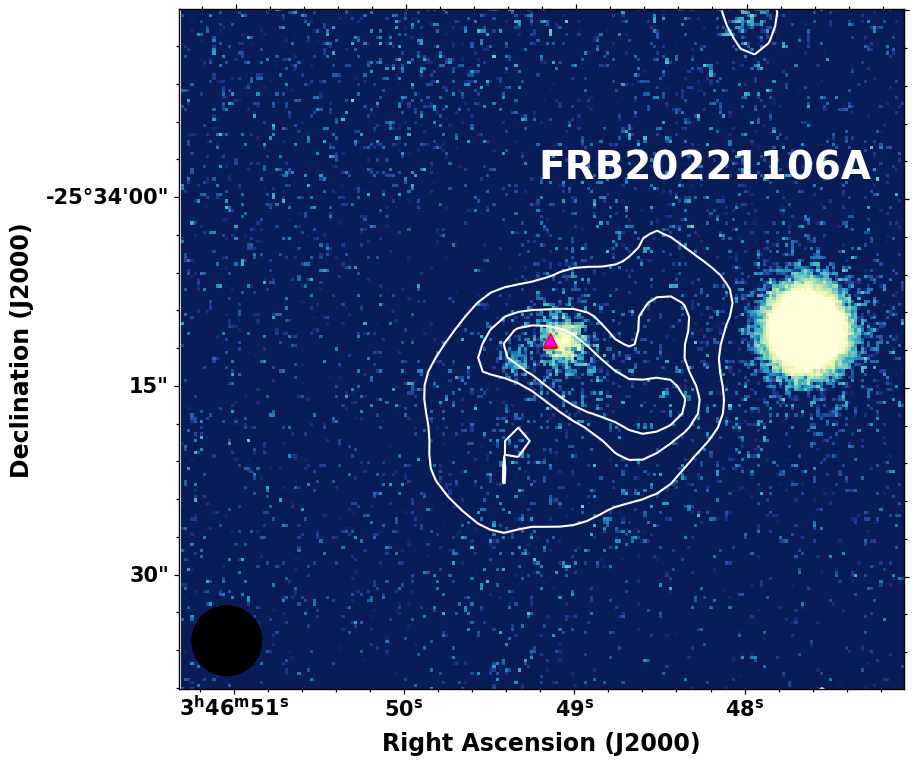 }
    \end{minipage}
\caption{Optical fields with the MeerKAT radio contours overlaid. The background is the Pan-STARRS $g$-band optical flux and the white contours represent the radio emission corresponding to $3,6,12,24$ times the rms of the image. The magenta triangle is the position of the FRB and black circle in the bottom left corner represents the size of the MeerKAT beam.}
    \label{fig:Optical-counterparts}
\end{figure*}

Two epochs were observed toward this FRB position and there is a detection in both. The host galaxy of FRB20190608B has been studied in some detail \citep{Chittidi_2021, Bhandari_2022}, thus the FRB is localised to a star-forming spiral galaxy that contains an AGN at its centre. We found an optical FITS image in the Pan-STARRS catalog (Figure~\ref{fig:Optical-counterparts}) indicating an optical source coinciding with the detected radio source. An X-ray FITS image was found in the \textit{Chandra} archive. There seems to be a bright X-ray excess spatially coinciding with the radio source position. It is plausible that the AGN at the centre of the host galaxy may be responsible for the radio and X-ray emission.

\subsubsection{FRB20191001A field}
\begin{figure*}
    \centering
    \begin{minipage}{1.0\textwidth}
    \centering
    \includegraphics[scale=0.31]{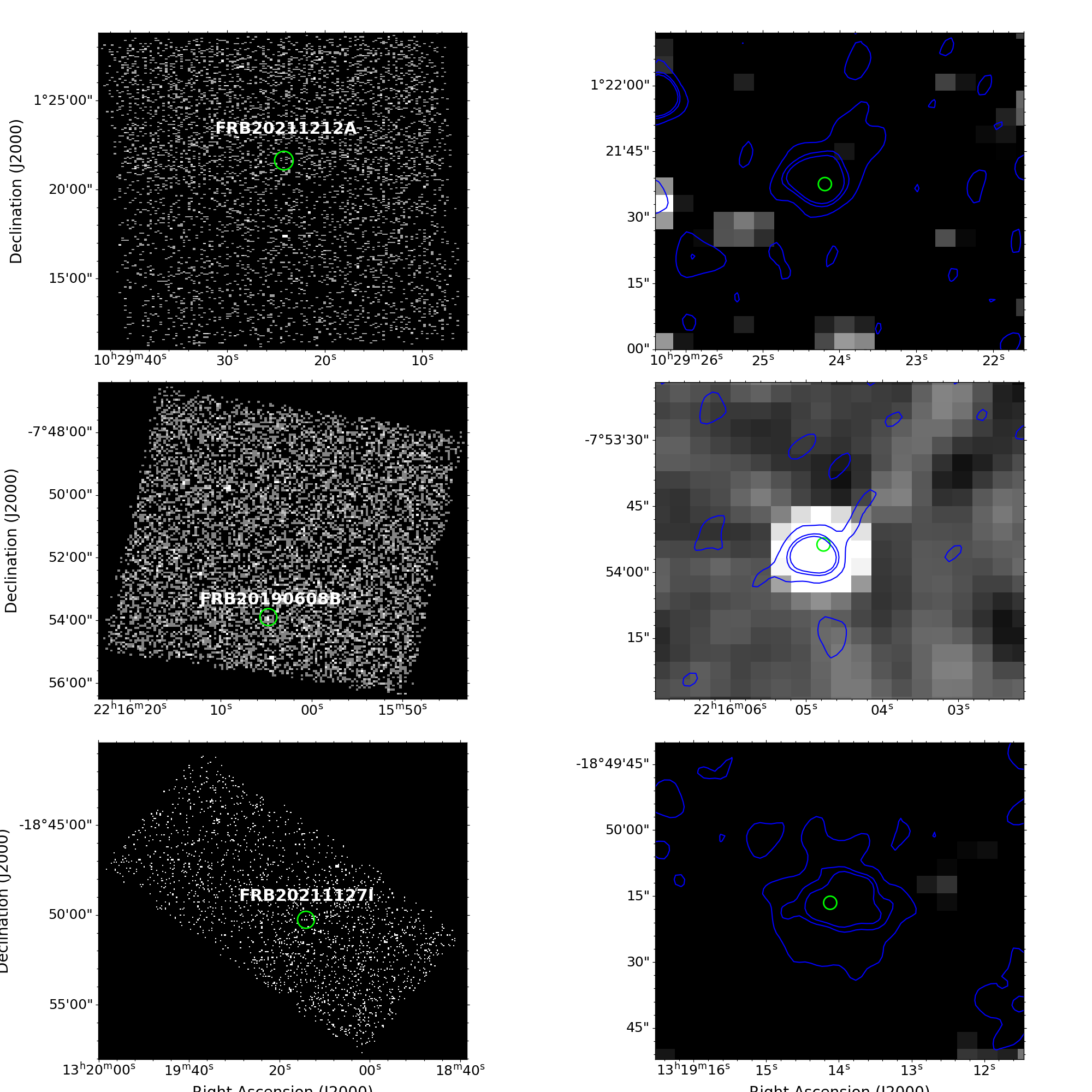}
    \caption{The \textit{Chandra} X-ray images coinciding with FRB positions. \textit{Left:} The green circle represents the FRB position.  \textit{Right:} The zoomed image close to the FRB position. The blue contours represent the radio emission corresponding with $3,6,12,24$ times the rms of the image, and the green circle is the FRB position.}
    \label{fig:chandra-xray}
\end{minipage}

\begin{minipage}{1.0\textwidth}
    \centering
    \centering
    \includegraphics[scale=0.37]{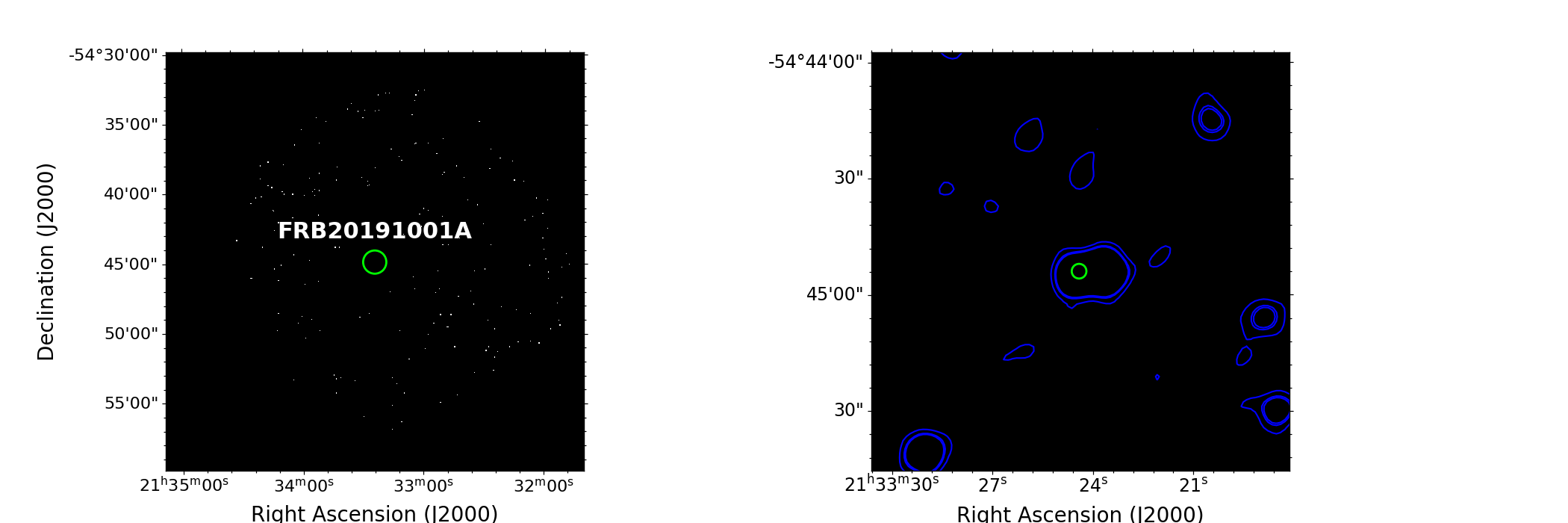}
    \caption{The \textit{Swift} X-ray images coinciding with FRB positions. \textit{Left:} The green circle represents the FRB position. \textit{Right:} The zoomed image close to the FRB position. The blue contours represent the radio emission corresponding with $3,6,12,24$ times the rms of the image, and the green circle is the FRB position.}
    \label{fig:swift-xray}
    \end{minipage}
\end{figure*}

The FRB20191001A position was observed for two epochs, and there is a detection in both epochs. The radio emission shows an interesting, extended morphology. \citet{yamanaka2024} used ALMA to spatially resolve this source into two components, which are two spiral galaxies with a possibility of them merging. The FRB is localised on the north of the left spiral galaxy (cf.\ Figure~\ref{fig:Optical-counterparts1}). The optical FITS image of these galaxies was found in the DESI legacy survey. The X-ray FITS image from the \textit{Swift} catalog, however, indicates that there is no detection at the FRB and host galaxy position (Figure~\ref{fig:swift-xray}). The detailed properties of a host galaxy are presented in \cite{Heintz2020}. \cite{Bhandari_2020a} used Australia Telescope Compact Array (ATCA) data to search for a compact PRS that is associated with the FRB. They detected low-level diffuse radio emission, however, they did not find a compact PRS above the flux density of $15 \mu$ Jy.

\subsubsection{FRB20200906A field}
This is a single-epoch observation. We found an optical FITS image in the Pan-STARRS catalog indicating an elongated source, and \citet{Bhandari_2022} reported it as a star-forming galaxy. However, no X-ray source coinciding with the radio emission was found in the \textit{Chandra} catalog. 

\subsubsection{FRB20210410D field}
Two epochs were observed at this FRB position, with a detection in both. The host galaxy of FRB20210410D has been studied by \citep{Caleb-2023}, where the FRB is localised to the southeast of the host galaxy. \cite{Caleb-2023} reported that persistent emission close to the FRB was not observed consistent with the position of the galaxy using MeerKAT. By visual inspection, the peak of the radio source is offset from the centre of the optical source found in the DESI legacy survey catalog. No X-ray FITS images were found in the archives.

\subsubsection{FRB20210912A field}
This is a single-epoch observation. \citep{Marnoch2023} did not find a host galaxy, despite deep optical ($R > 26.7$ mag) and infrared (to $K_{\rm s} > 24.9$ mag) follow-up with the European Southern Observatory's Very Large Telescope (ESO VLT). We found no optical source spatially coinciding with the detected continuum radio emission. Thus, there is no redshift information on the FRB and the radio source. The radio emission could be emitted by a source that is highly redshifted or it could be an obscured galaxy at the optical wavelength.

\subsubsection{FRB20210807D field}
This is a single-epoch observation at the FRB position. An optical FITS image from the DESI-legacy-survey reveals a spatially-coincident source with the radio emission. \cite{gordon2023} reported it as a quiescent galaxy. We did not find an X-ray source in the archival catalogs coinciding with the detected radio source.

\subsubsection{FRB20211127I field}
This is a single-epoch observation, and the optical and X-ray FITS images spatially coinciding with the radio emission's position were found in Pan-STARRS and \textit{Chandra}, respectively. The optical counterpart shows a spiral galaxy that was reported to be a star-forming galaxy by \citet{gordon2023}.  

\subsubsection{FRB20211212A field}
This is a single-epoch observation. We found an optical source that spatially coincides with the radio emission in the Pan-STARRS catalog. There does not seem to be a significant X-ray excess from the \textit{Chandra} catalog coinciding with the radio emission. \citet{gordon2023} studied the host galaxy of this FRB, presenting the photometric and spectrometric observations of the host.

\subsubsection{FRB20210320C field}
This is a single-epoch observation. An optical FITS image was found in Pan-STARRS. \citet{gordon2023} reported that it is a star-forming galaxy. No X-ray source coinciding with this position was found in the archival catalogs.

\subsubsection{FRB20220501C field}
This FRB position was observed for three epochs. There is no detection in the first epoch, however, there is a detection in the second and third epochs, and this suggests flux variability. The variability of these three epochs is of day-timescales, since we observed this field on 9, 25, and 26 November (see Table~\ref{tab:info1} and~\ref{tab:info2}). The flux significantly decreases (the radio source is 9 times fainter in third epoch compared to the second epoch) as though the source is fading away. The variability could be as a result of intrinsic variability or it could be the  result of scintillation effects. Therefore, more data are required to study this source.

There is no detailed study of the host (optical source) of FRB20220501C in the literature because it is not catalogued. However, there is a faint optical source from the DESI legacy survey spatially coinciding with the detected radio emission and FRB position and there is a bright ($V = 11.6$ magnitude) star located $15^{''}$ from this optical source. The Set of Identifications, Measurements, and Bibliography for Astronomical Data (SIMBAD) database (https://simbad.u-strasbg.fr/simbad/sim-fcoo) did not identify nor classify the optical source. No X-ray FITS images were found in the archival catalogues that are spatially associated with the radio source. A detailed study on the host of this FRB is encouraged.

\subsubsection{FRB20220725A field}
This is a single-epoch observation, and a spatially-coincident optical source was found in DESI-legacy-survey. \citet{Shannon2025} claimed that the FRB coincides with the catalogued galaxy WISEA J233315.68-355925.0 \citep{1990Maddox}, with an optical magnitude of $b_J = 19.0$. No coinciding X-ray image was found. 

\subsubsection{FRB20220918A field}
This is a single-epoch observation, with the radio source exhibiting an interesting morphology. This source has a tail (which may perhaps be ejected material from the centre). A high-resolution observation may reveal multiple sources/clumps in the region. There is no detailed study on the host of FRB20220918A in the literature. An optical FITS image was found in DESI legacy survey, and SIMBAD classifies the optical source as the Small Magellanic Cloud (SMC). This raises an intriguing possibility that the radio continuum emission may perhaps be associated with the SMC. Interestingly, the FRB is at the centre of both the radio and optical sources. We could not find X-ray FITS images in the archives coinciding with the radio emission. 

\subsubsection{FRB20221106A field}
\begin{figure*}
    \centering
    \includegraphics[width=1.0\linewidth]{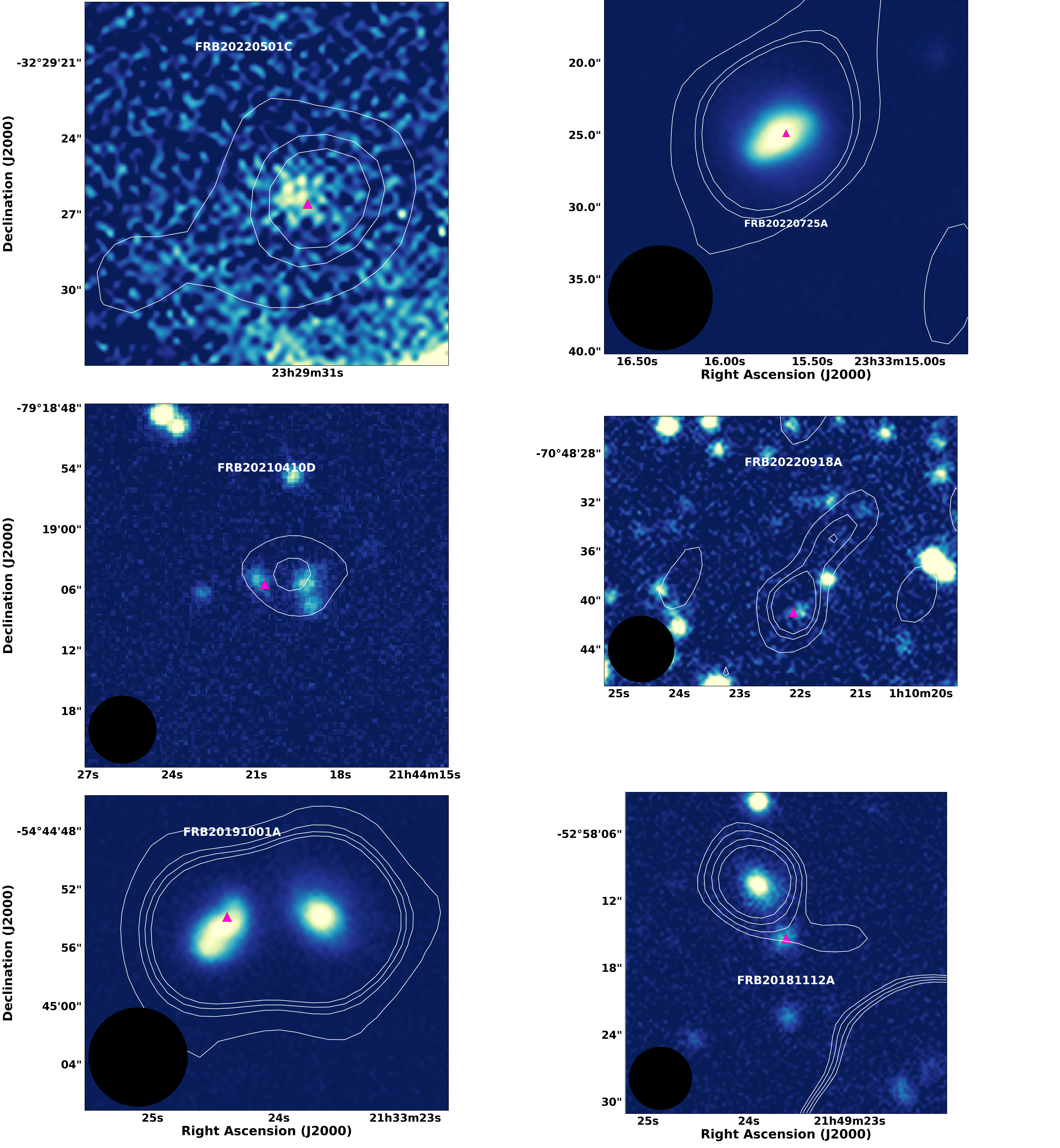}
    \caption{Optical fields with the MeerKAT radio contours overlaid. The background is the DESI-Legacy survey $ grz$-band optical flux and the white contours represent the MeerKAT radio emission corresponding to $3,6,12,24$ times the rms of the image. The magenta triangle is the position of the FRB and black circle in the bottom left corner represents the size of the MeerKAT beam.}
    \label{fig:Optical-counterparts1}    
\end{figure*}

The FRB20221106A position was observed for three epochs, and there is a detection in all three. There is a hint of variability in the source morphology and peak flux. We found an optical source spatially coinciding with the radio emission (with high peak flux, from the second epoch) in the DESI legacy survey catalogue (Figure~\ref{fig:Optical-counterparts1}). \citet{Shannon2025} claimed that the galaxy WISEA J034649.07-253411.7 with an optical magnitude of $b_J = 19.5$ \citep{1990Maddox} coincides with the FRB position. There are no other recent studies on this host galaxy.  Therefore, the host galaxy cannot be identified. SIMBAD suggests that it is a star cluster. The peak of the radio source is offset from the centre of the optical source. No X-ray FITS images were found in the archival catalogues. 

\begin{table*}
\centering
 \caption{List of FRB name, dispersion measure (DM), the redshift and where it is obtained and also the positions in right ascension and declination taken from ASKAP/CRAFT and MeerTRAP. We also indicate which is a repeating source.}
 \label{tab:appendix}

 \begin{tabular}{cccccccc}
  \hline
  Source name & Ref. & DM & $z$     & RA & DEC & Repeater? & Instrument\\
              & & pc/cm$^{-3}$ &    &     & &\\ 
    \hline
    FRB20181112A & \citep{Prochaska2019}& 589.0 & 0.4755 & 21:49:23.63 & –52:58:15.4 & N & ASKAP\\
    \hline
    FRB20190102C & \citep{Macquart2020}& 364.5 & 0.2912 & 21:29:39.76 & –79:28:32.5 & N & ASKAP\\
    \hline
    FRB20190608B & \citep{Macquart2020} & 339.5 & 0.1178 & 22:16:04.77 & –07:53:53.7 & N & ASKAP\\
    \hline
    FRB20190611B &\citep{Macquart2020} & 322.2 & 0.3778 & 21:22:58.94 & –79:23:51.3 & N & ASKAP\\
    \hline
    FRB20191001A &\citep{Bhandari_2020a} & 506.92 & 0.234 & 21:33:24.37 & -54:44:51.4 & N & ASKAP\\
    \hline
    FRB20191228A &\citep{Bhandari_2022} & 297.5 & 0.2432 & 22:57:43.33 & –29:35:38.8 & N & ASKAP\\
    \hline
    FRB20200430A &\citep{Heintz2020} & 380.1& 0.1608 & 15:18:49.55 & +12:22:34.8 & N & ASKAP\\
    \hline
    FRB20200906A & \citep{Bhandari_2022}& 577.8 & 0.3688 & 3:33:58.94 & -14:04:59.9 & N & ASKAP\\
    \hline
    FRB20210117A &\citep{Bhandari_2023} & 730 & 0.2145 & 22:39:55.015 & -16:09:05.45 & N & ASKAP\\
    \hline
    FRB20210320C & \citep{james2022, gordon2023}& 384.8 & 0.2797 & 13:37:50.09 & -16:07:21.7 & N & ASKAP\\
    \hline
    FRB20210410D & \citep{Caleb-2023}& 578.8 &0.1415 & 21:44:20.63 & -79:19:05.6 & N & MeerTRAP\\
    \hline
    FRB20210807D & \citep{james2022, gordon2023}& 251.9 & 0.1293 & 19:56:53.14 & -0:45:44.5 & N & ASKAP\\
     \hline
     FRB20210912A & \citep{Marnoch2023}& 1234.5 & -- & 23:23:10.44 & –30:24:20.1 & N & ASKAP\\
     \hline
    FRB20211127I &\citep{james2022, gordon2023} &234.83 & 0.0469 & 3:19:14.08 & 18:50:16.7 & N & ASKAP\\
    \hline
    FRB20211212A &\citep{james2022, gordon2023} & 206 & 0.0707 & 10:29:24.16 & 1:21:37.7 & N & ASKAP\\
    \hline
    FRB20220105A &\citep{gordon2023} & 583 & 0.2785 & 13:55:12.81 & +22:27:58.4 & N & ASKAP\\
    \hline
    FRB20220222C & \citep{pastor2025}& 1071.2 &0.853 & 13:35:37.06 & -28:01:37.2 & N & MeerTRAP\\
    \hline
    FRB20220330 & -- & 900 & -- & 18:17:52.83 & -24:23:42.5 & N & MeerTRAP\\
    \hline
    FRB20220501C & \citep{Shannon2025}& 449.5 & 0.381 & 23:29:31.00 & –32:29:26.6 & N & ASKAP\\
    \hline
    FRB20220717A &\citep{rajwade2024} &637 & 0.36295& 19:33:13.0 & -19:17:15.8 & N & MeerTRAP\\
    \hline
    FRB20220725A &\citep{Shannon2025} & 290.4 & 0.1926 & 23:33:15.65 & –35:59:24.9 & N & ASKAP\\ 
    \hline
    FRB20220905A &\citep{rajwade2024} &800.6 & --& 16:54:19.79 & -20:04:16.2 & N & MeerTRAP\\
    \hline
    FRB20220918A &\citep{Shannon2025} & 656.8 & 0.491 & 01:10:22.11 & –70:48:41.0 & N & ASKAP\\
    \hline
    FRB20221106A &\citep{Shannon2025} & 343.8 & 0.2044 & 3:46:49.09 & -25:34:10.5 & N & ASKAP\\
     \hline
   FRB20230125D & \citep{pastor2025}& 640.08& 0.3265& 10:00:49.33 & -31:32:41.1 & N & MeerTRAP\\
    \hline
   \end{tabular}%}
    \label{tab:my_label}
\end{table*}

%%%%%%%%%%%%%%%%%%%%%%%%%%%%%%%%%%%%%%%%%%%%%%%%%%

% Don't change these lines
\bsp	% typesetting comment
\label{lastpage}
\end{document}